\documentclass{ieeeaccess}
\usepackage{epsf,graphicx}
\usepackage{amsmath,amssymb,amsfonts}
\usepackage{subfigure} 
\graphicspath{{./figs/}}
\usepackage{float}
\usepackage{multirow}
\usepackage{multicol}
\usepackage{cleveref}
\usepackage{textcomp}
\usepackage{cite}
\usepackage{algorithmic}

\usepackage{bm}
\makeatletter
\AtBeginDocument{\DeclareMathVersion{bold}
\SetSymbolFont{operators}{bold}{T1}{times}{b}{n}
\SetSymbolFont{NewLetters}{bold}{T1}{times}{b}{it}
\SetMathAlphabet{\mathrm}{bold}{T1}{times}{b}{n}
\SetMathAlphabet{\mathit}{bold}{T1}{times}{b}{it}
\SetMathAlphabet{\mathbf}{bold}{T1}{times}{b}{n}
\SetMathAlphabet{\mathtt}{bold}{OT1}{pcr}{b}{n}
\SetSymbolFont{symbols}{bold}{OMS}{cmsy}{b}{n}
\renewcommand\boldmath{\@nomath\boldmath\mathversion{bold}}}
\makeatother

\def\BibTeX{{\rm B\kern-.05em{\sc i\kern-.025em b}\kern-.08em
    T\kern-.1667em\lower.7ex\hbox{E}\kern-.125emX}}

\begin{document}
\history{Date of publication xxxx 00, 0000, date of current version xxxx 00, 0000.}
\doi{10.1109/ACCESS.2024.0429000}

\title{Unbalanced CRLH Leaky-wave Antenna With Broadside Radiation Based On Spin Photonic Topological Insulator Featured Hexagonal Configuration In Armchair Arrangement}
\author{SAYYED AHMAD ABTAHI\authorrefmark{1},
        MOHSEN MADDAHALI\authorrefmark{1},
        AND AHMAD BAKHTAFROUZ\authorrefmark{1}}
\address[1]{Department of Electrical and Computer Engineering,
        Isfahan University of Technology,
        Isfahan 84156-83111, Iran }
\markboth
{S. A. Abtahi \headeretal: Realizable Leaky Wave Antenna based on Spin Photonic Topological Insulators}
{S. A. Abtahi \headeretal: Realizable Leaky Wave Antenna based on Spin Photonic Topological Insulators}
\corresp{Corresponding author: M. Maddahali (e-mail: maddahali@iut.ac.ir).}

\begin{abstract}
A new X-band leaky-wave antenna has been developed using spin photonic topological insulators. This antenna features a hexagonal unit cell arranged in an armchair configuration. This arrangement provides advantages over the zigzag configuration, particularly by offering a wider operational region and a more suitable pattern.
To design the structure, a parametric study on the cell dimensions has been conducted, and another study has designed the transition region to couple the topological structure with the classical line.
The proposed antenna has a low profile and illuminates two sides of the structure simultaneously. Additionally, it offers a 53-degree scanning range and a bandwidth of 2.7 GHz, making it a groundbreaking improvement over other leaky-wave antennas that utilize photonic topological insulators.
The proposed antenna is an unbalanced CRLH leaky-wave antenna capable of radiating in both backward and forward directions. Notably, as it transitions through the broadside within its scanning range, there is no significant drop in performance, even in the presence of an open stop band. To the best of our knowledge, this characteristic is unique among unbalanced CRLH leaky-wave antennas.
\end{abstract}


\titlepgskip=-21pt

\maketitle

\section{Introduction}

\PARstart{M}{etasurfaces} are the two-dimensional counterparts of bulky metamaterials, composed of subwavelength unit cells that exhibit unconventional electromagnetic properties\cite{ref6}.
These low-profile structures are a subject of extensive research\cite{caloz2005,achouri2021meta,balanis2024EM} and have a wide range of applications, including absorption, beam shaping, and radiation\cite{chen2016metasurfaceISO,hsiao2017metasurface,asadchy2018bianisotropic,li2018metasurfaces,bukhari2019metasurfaces}.

 For guiding applications, the modes of the structure should be situated in the slow wave region. Conversely, for radiating applications, the propagating modes need to be positioned in the fast wave region\cite{ref13}. 
The realized leaky-wave antennas (LWAs) using metasurfaces in 1D\cite{grbic2011,memarian,1D-Leaky} and 2D\cite{bakhtafrouz2015,minattiflat,minattispiral} dimensions have demonstrated versatility in achieving low-profile multi-beam configurations\cite{ref14}, arbitrary shapes\cite{minattiIsoflux}, and wide scanning ranges\cite{wideScanning}.
A new application of metasurfaces enables propagation along a line between two impedance surfaces that support TE/TM polarization, which is referred to as line waves\cite{bisharat2017, KhavasiLinewave, Maci_line, khodadadidual, xulinewave}. These structures offer an ultra-wide operating bandwidth; however, they are prone to scattering when encountering sharp turns or structural defects\cite{ref4}.

Photonic topological insulators (PTIs) are a recent development designed for use in guiding-wave and antenna applications\cite{khanikaev2024}. They are robust against scattering caused by manufacturing defects and sharp turns\cite{ozawa}. One type of PTI is known as spin PTIs, which preserve time-reversal symmetry and are realized using spin-degenerate metamaterials\cite{khanikaev2013}.
Spin PTI metasurfaces are introduced using a complementary hexagonal unit cell\cite{bisharat2019}, as illustrated in Fig. \ref{HexCell}. This unit cell exhibits fourfold degeneracy at the symmetry points of K/K’ in its irreducible Brillouin zone (IRBZ). By combining the complementary parts, strong effective magnetoelectric coupling results in a nontrivial band gap at the K/K’ points. The topological structure is formed by aligning the unit cell with its flipped version, creating edge modes at the interface. The unidirectional propagation of line waves at the interface of two structural components is investigated in zig-zag and armchair configurations. 

Using PTIs for leaky-wave antenna applications is not limited to spin PTIs\cite{chernAntenna,chernantennaPeriodic}. However, such structures require YIG photonic crystals and a static magnetic field bias, which limits their practicality. In contrast, there is no need for such requirements in spin PTI metasurfaces.

A leaky-wave antenna based on spin PTI and utilizing a zigzag arrangement has been investigated in \cite{SinghAntenna}. However, this approach presents several significant challenges, including a narrow bandwidth, limited backward radiation scanning, coupling between radiating elements, and an inadequate radiation pattern.

A proven solution to these issues involves transforming the unit cell from a hexagonal configuration to a 30-degree rhombic shape. This modification promotes straight-line propagation, which effectively reduces coupling between radiating elements \cite{abtahi2}. While the benefits of using 30-degree rhombic cells are considerable, it is important to note that the radiation direction is limited to backward radiation, and the scanning range is restricted to 22.5 degrees.


This paper proposes that adopting an armchair arrangement, instead of a zigzag design, effectively addresses the identified issues and facilitates backward-to-forward beam scanning. A crucial point is that broadside radiation in unbalanced composite right/left-handed (CRLH) leaky-wave antennas presents significant challenges, often leading to a sudden drop in gain due to an open stopband. However, in the proposed antenna with the armchair arrangement, the impact of this phenomenon is significantly reduced. Simulations were conducted using Ansys HFSS and were subsequently repeated in CST Studio to validate the results.


   
\section{Motivation to use armchair arrangement}
\begin{figure}[t!]
	\begin{center} 
	\subfigure[]{
	\includegraphics[width=0.45\textwidth]{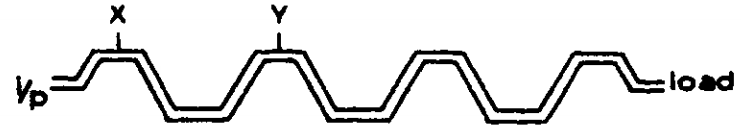}
	\label{trapz}
	}
	\hspace{0.005cm}
	\subfigure[]{
	\includegraphics[width=0.45\textwidth]{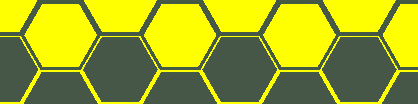}
	\label{armchair}
	}
	\end{center}
\caption{\subref{trapz} The trapezoidal serpent line by microstrip\cite{microstrip_james} \subref{armchair} hexagonal unit cell of \cite{bisharat2019} in armchair arrangement. }
\label{ArmTrapz}
\end{figure}

The concept of utilizing the armchair arrangement as a leaky-wave antenna derives directly from the trapezoidal serpent line microstrip used in antenna applications\cite{microstrip_james}. As illustrated in Fig.\ref{ArmTrapz}, there is a clear analogy between these two structures.
Initially, to compare the zigzag and armchair configurations, the dimensions were selected to match those in \cite{SinghAntenna}. Specifically, a periodic length of 7 mm and a border width of 0.25 mm were used. The substrate employed is Rogers/Duroid 5880, which has a thickness of 0.127 mm, a permittivity of 2.2, and a tangent loss factor of 0.0009. It is important to note that these dimensions are not final and may vary depending on the specific application.
The parametric study on period length and border width identifies the dimensions necessary for designing an X-band leaky-wave.

\begin{figure}[h!]
	\begin{center}
	\subfigure[]{
	\includegraphics[width=0.47\textwidth]{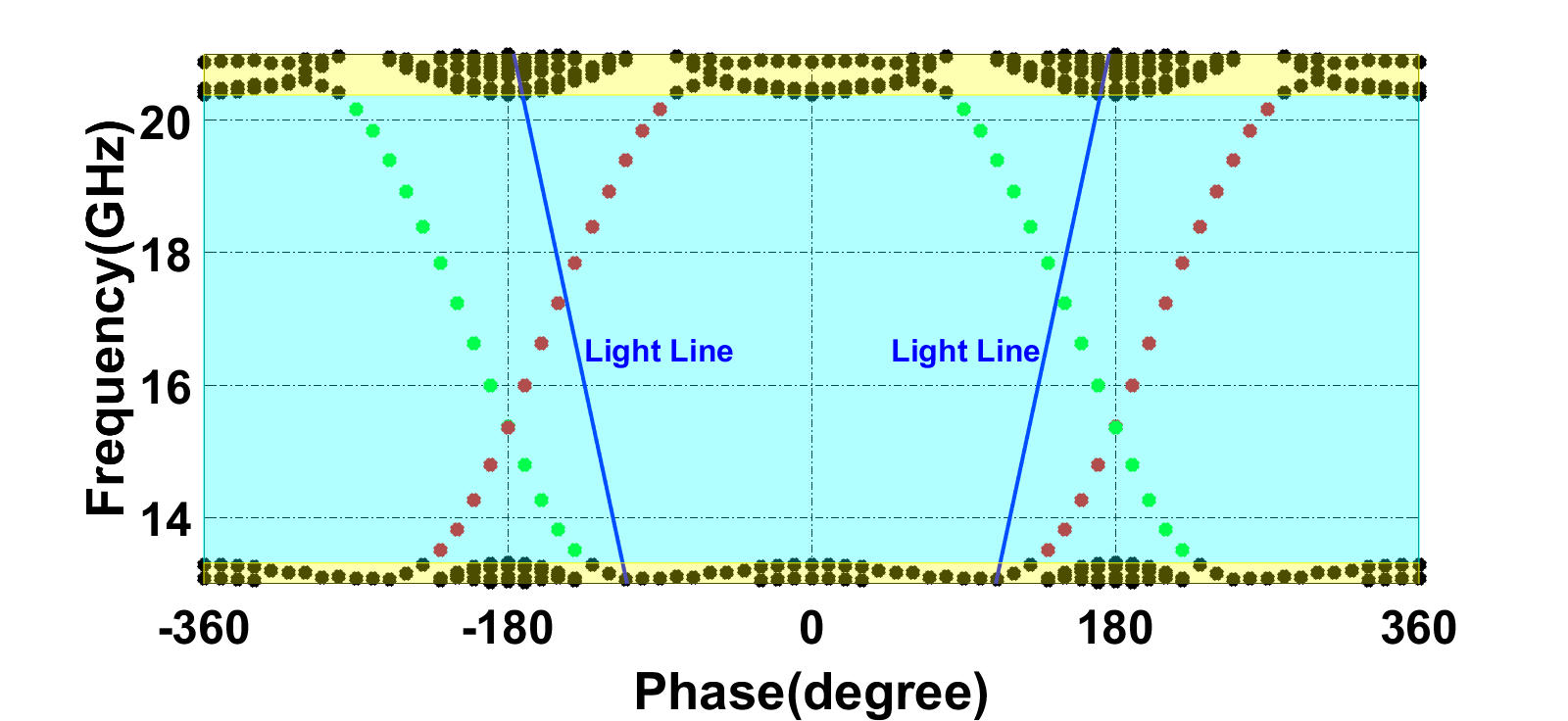}
	\label{Hex_Ribbon_Zigzag}
	}
	\hspace{0.005cm}
	\subfigure[]{
	\includegraphics[width=0.47\textwidth]{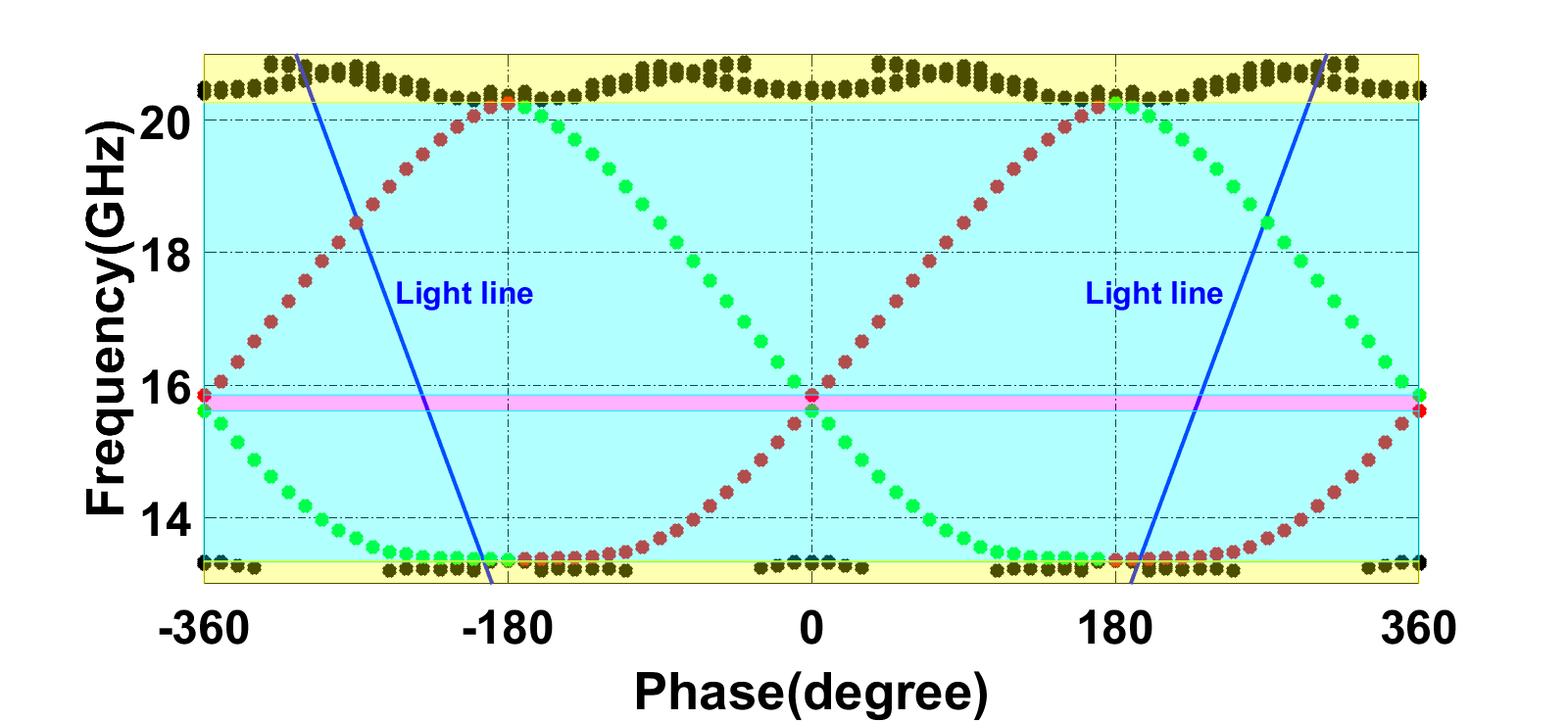}
	\label{Hex_Ribbon_Armchair}
	}
	\end{center}
\caption{ 
The dispersion diagrams for a single ribbon of the structure are illustrated for both the zigzag arrangement \subref{Hex_Ribbon_Zigzag} and the armchair arrangement \subref{Hex_Ribbon_Armchair}. In these diagrams, the yellow region marked with black dots represents the bulk modes, while the cyan region, indicated by red and green dots, represents the edge modes. The red and green dots are used to distinguish between two pseudospins. Additionally, the magenta region in \subref{Hex_Ribbon_Armchair} indicates the presence of a bandgap for the armchair arrangement.
}
\label{Ribbon}
\end{figure}

The dispersion diagram for one ribbon from both arrangements is calculated using Ansys HFSS, as shown in Fig.\ref{Ribbon}. The primary differences are the alteration in group velocity with increasing frequency and the location of the light line in both diagrams. The group velocity transitions from positive to negative in the zigzag arrangement, whereas it behaves oppositely in the armchair configuration.
In Fig. \ref{Hex_Ribbon_Zigzag}, most of the edge modes are situated in the slow-wave region, rather than the fast-wave region. In contrast, for the armchair arrangement, most of the edge modes are located within the fast-wave region. Consequently, the armchair structure radiates across the entire topological bandgap, while the zigzag structure only radiates above 17 GHz. A key feature of the zigzag arrangement is the presence of a Dirac point in the dispersion diagram, which is absent in the armchair configuration. Therefore, the existence of a bandgap in Fig. \ref{Hex_Ribbon_Armchair} is not an unprecedented occurrence \cite{graphene_review}.

\subsection{Parametric Study}
\begin{figure}[h!]
	\begin{center}
	\subfigure[]{
			\includegraphics[width=0.48\textwidth]{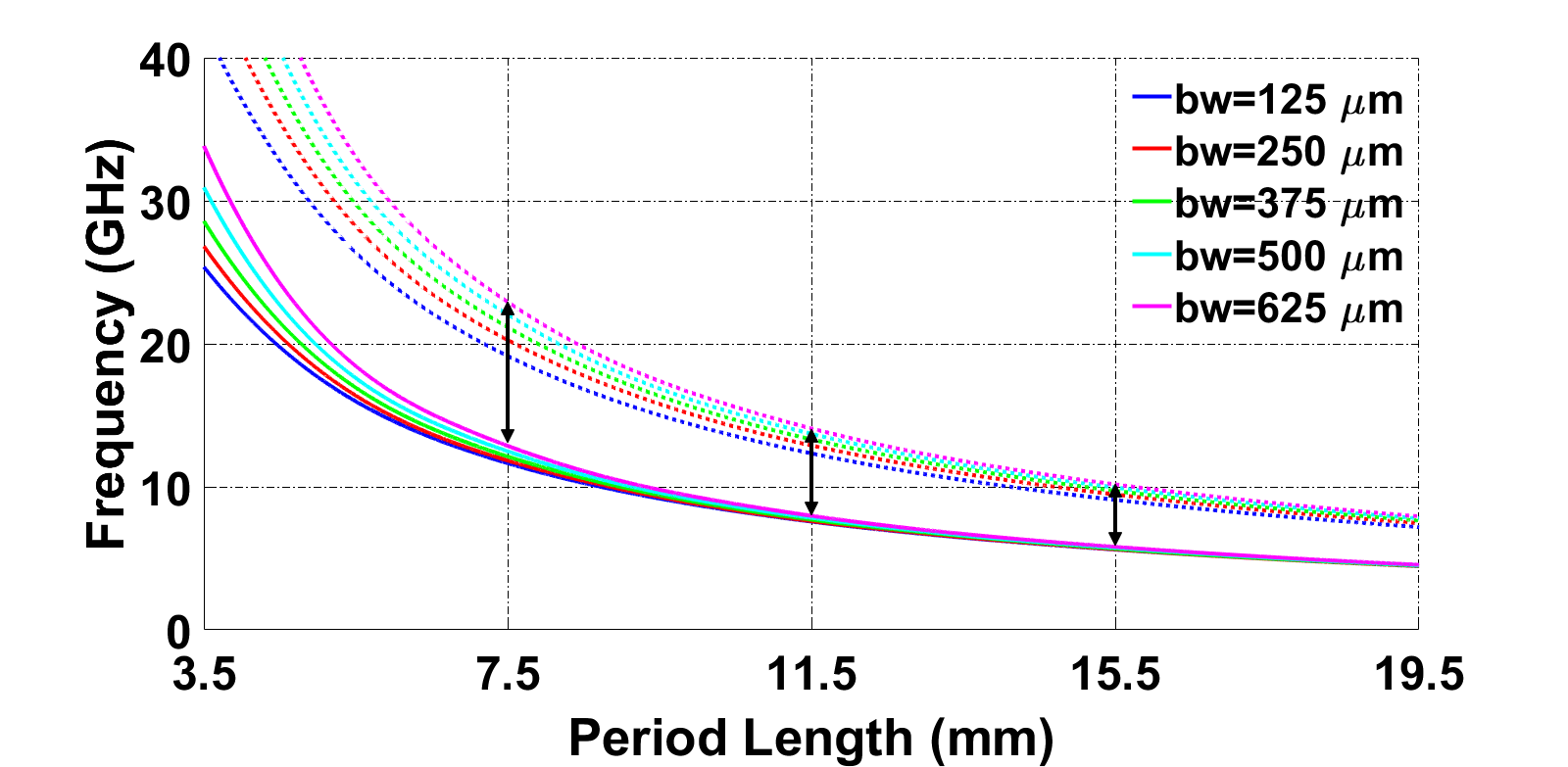}
			\label{parameter_study_thickness_0127_Max_Min_Bandgap}	
				}
	\hspace{0.0005cm}
	\subfigure[]{
			\includegraphics[width=0.48\textwidth]{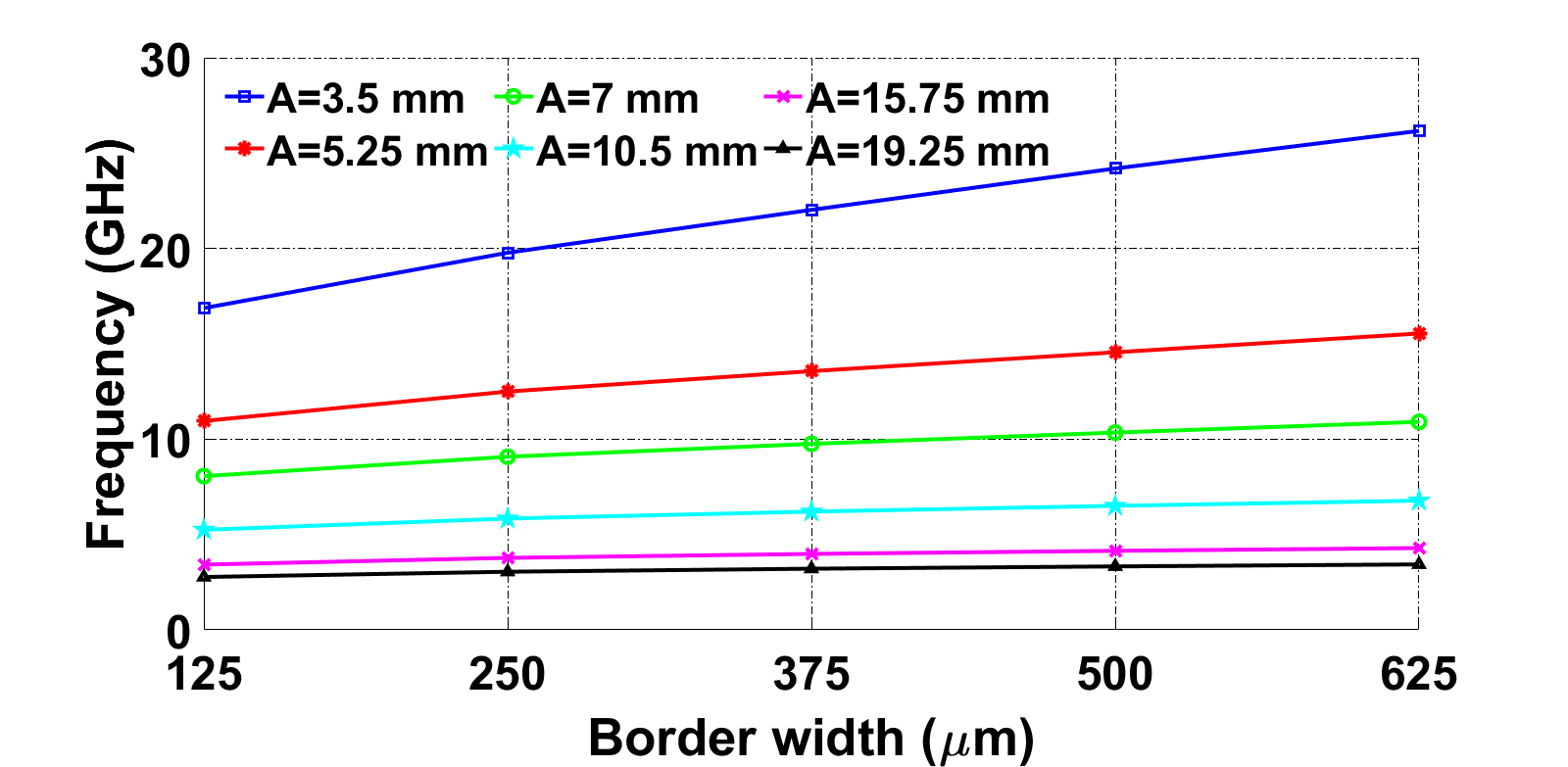}
			\label{parameter_study_thickness_0127}	
	}

	\end{center}
	\caption{\subref{parameter_study_thickness_0127_Max_Min_Bandgap} Variation of maximum of first degenerate band ( solid lines ) and minimum of the second degenerate band ( doted line ) by changing the period length for different border widths (bw). The arrows shows the decline of bandgap by increasing the period length.\subref{parameter_study_thickness_0127} Variation of the bandgap width by increasing the border width for some period length (A).}
	\label{parameter_study}
\end{figure}

The objective was to design an X-band leaky-wave antenna using a spin topological insulator featuring a hexagonal lattice arranged in an armchair configuration. Therefore, it is essential to analyze the cell parameter to determine the appropriate cell periodicity and border width.
The Fig.\ref{parameter_study} illustrates how the limits of the bandgap vary with changes in the period length and border width. Specifically, Fig. \ref{parameter_study_thickness_0127_Max_Min_Bandgap} demonstrates that both the maximum and minimum bandgap will decrease as the period length increases. Furthermore, the decline rate of the upper limit (the minimum of the second degenerate band) is greater than that of the lower limit (the maximum of the first degenerate band). 
On the other hand, Fig.\ref{parameter_study_thickness_0127} shows that the increasing of the border width will lead to a rise in the bandgap width, with a more significant rate of increase observed for shorter period lengths.

Based on the parametric study, a period length of 12 mm and a border width of 500 µm were selected to optimize the operational bandwidth within the X-band frequency range.

\begin{figure}[h!]
	\begin{center}
	\subfigure[]{
			\includegraphics[width=0.35\textwidth]{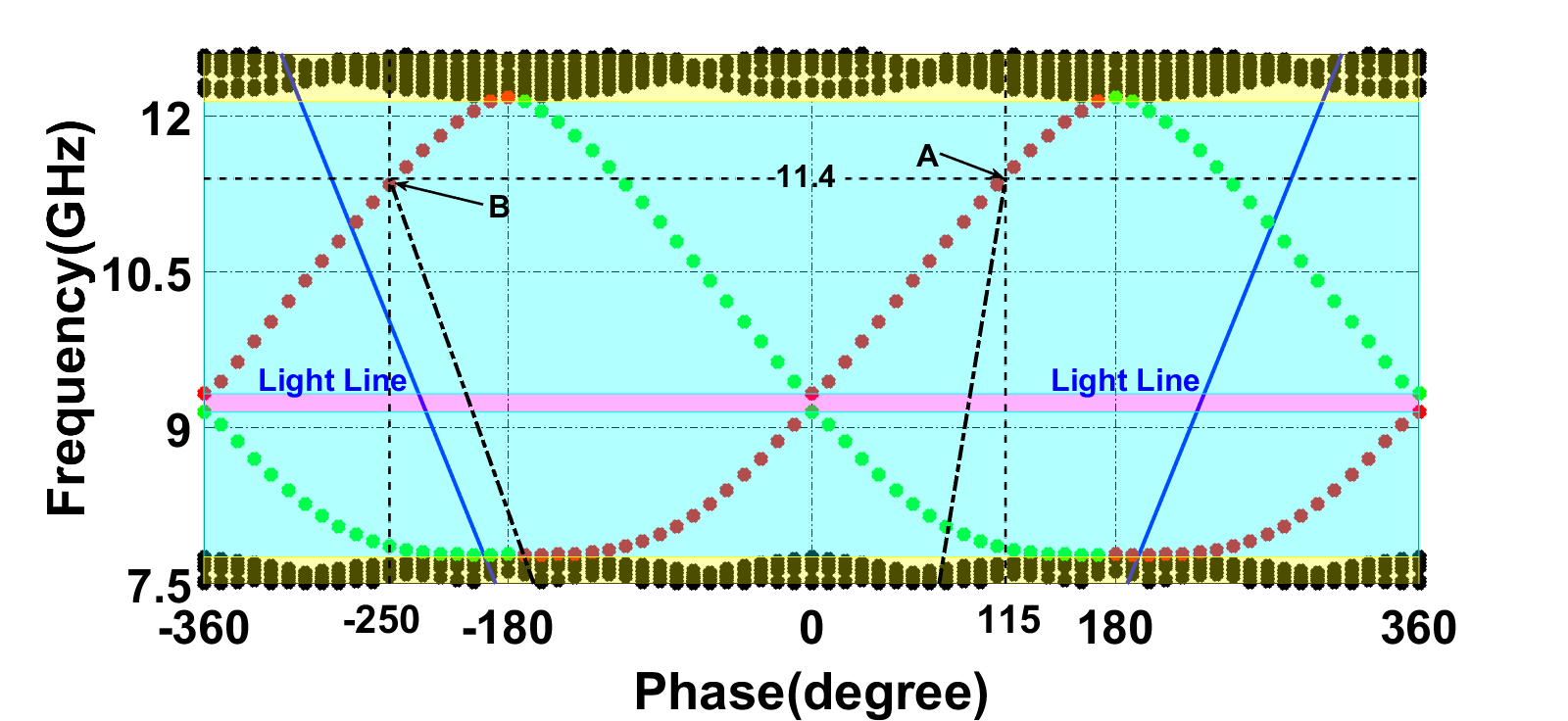}
			\label{Armchair_Ribbon_dispersion_a_12_b_05}	
				}
	\hspace{0.0005cm}
	\subfigure[]{
			\includegraphics[width=0.07\textwidth]{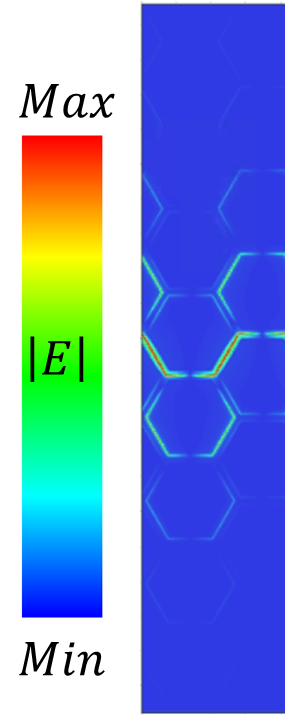}
			\label{a=12_b=0.5_Armchair}	
	}

	\end{center}
	\caption{ 
	\subref{Armchair_Ribbon_dispersion_a_12_b_05}The dispersion diagram of a hexagonally structured ribbon arranged in an armchair configuration is illustrated, demonstrating the presence of edge modes in the X-band domain. This finding supports the results presented in the parameter study in Fig. \ref{parameter_study}. The red and green dots represent the edge modes, while the black dots indicate the bulk modes. Additionally, the magenta area indicates the limits of the bandgap. 
	\subref{a=12_b=0.5_Armchair}The mode profile of the edge mode indicates propagation as a linear wave at the center of the ribbon.
	}
	\label{Selected_Ribbon}
\end{figure}

To investigate the existence of edge modes and validate the selections, we calculated and plotted the dispersion diagram of a ribbon, as shown in Fig.\ref{Selected_Ribbon}. As anticipated, the bandgap encompasses the X-band frequency range, while the edge mode extends across the bandgap, except for the inherent bandgap present in the armchair arrangement.


In the lower frequencies of the topological bandgap, the edge mode exhibits left-handed behavior. As the frequency increases, this behavior transitions to right-handed. This right-handed behavior persists until the edge mode intersects with the light line. Beyond this frequency, two modes propagate with opposite phase velocities.

For instance, at 11.4 GHz, there are two propagating modes, indicated by points A and B. The slope of the lines from the origin to points A and B (represented by the black dot-dash line in Fig.\ref{Armchair_Ribbon_dispersion_a_12_b_05}) determines the phase velocity at each point, which has opposite signs. As the frequency increases, the absolute values of phase velocities become closer, but the rate of this change is more significant for the phase velocity of point B.

\subsection{Matching ASL with Topological structure}

The method introduced in \cite{davisASL} for the zig-zag arrangement cannot be applied to the armchair configuration, necessitating a repetition of the steps to design the transition region. To achieve proper coupling, the field profile of the topological structure must align with that of the ASL configuration. 

\begin{figure}[h!]
	\begin{center}
			\includegraphics[width=0.47\textwidth]{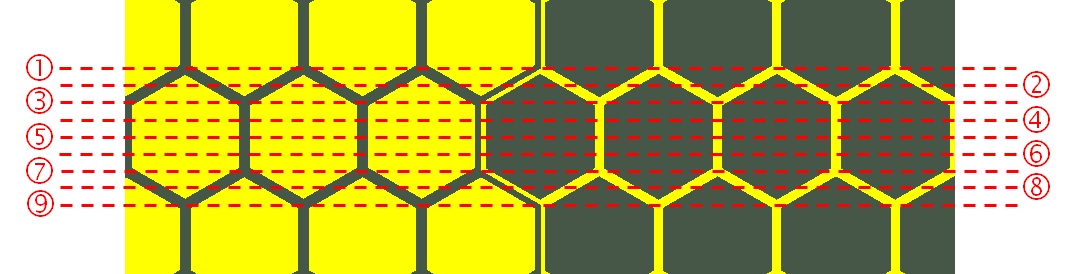}
	\end{center}
	\caption{The cut planes for investigating matching field profiles have been identified with nine dashed lines. 
	}
	\label{raw_armchair_cutplane}
\end{figure}

To determine the optimal coupling point between the structures, the field profile should be analyzed at each cut plane illustrated in Fig. \ref{raw_armchair_cutplane}, which are perpendicular to the propagation surface. The rotated versions of these cut planes are not considered due to the increase in the substrate area.

\begin{figure}[h!]
	\begin{center}
	\subfigure[]{
			\includegraphics[width=0.45\textwidth]{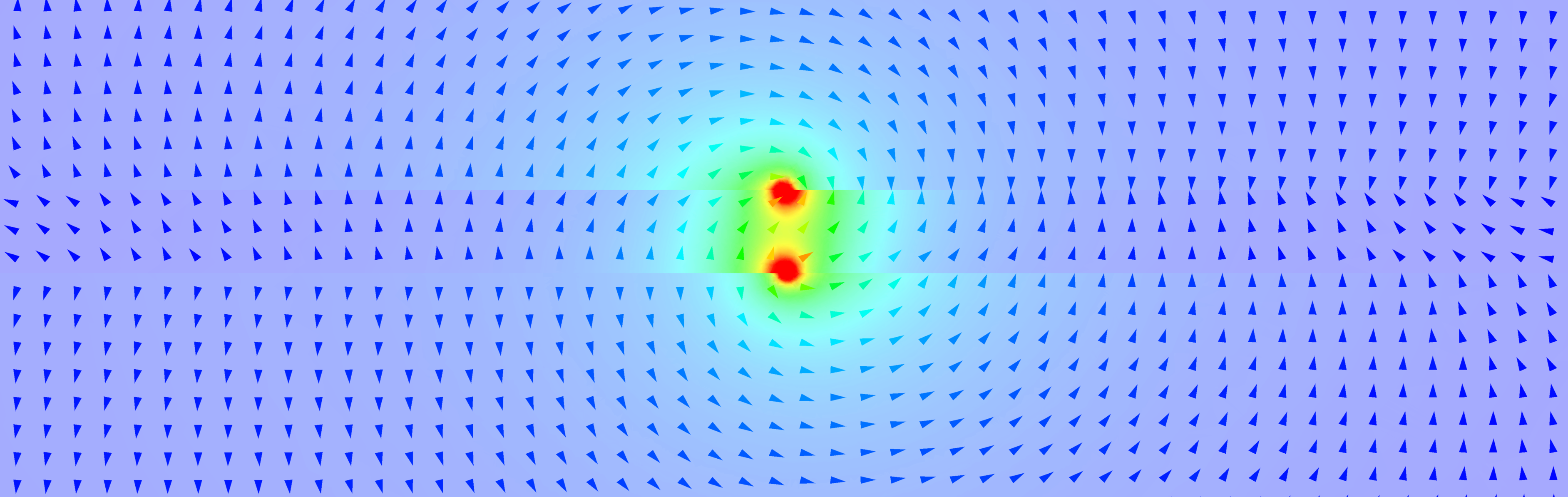}
			\label{ASL_S}	
				}
	\hspace{10cm}
	\subfigure[]{
			\includegraphics[width=0.13\textwidth]{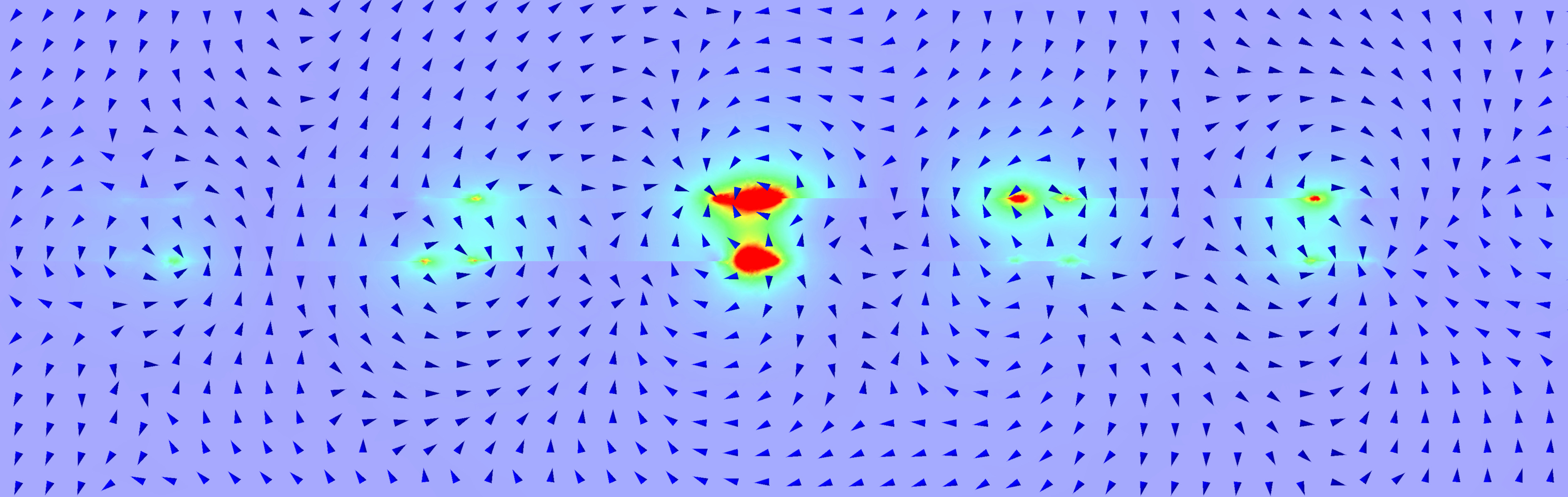}
			\label{Cut_1_S}	
	}
	\hspace{0.0005cm}
	\subfigure[]{
			\includegraphics[width=0.13\textwidth]{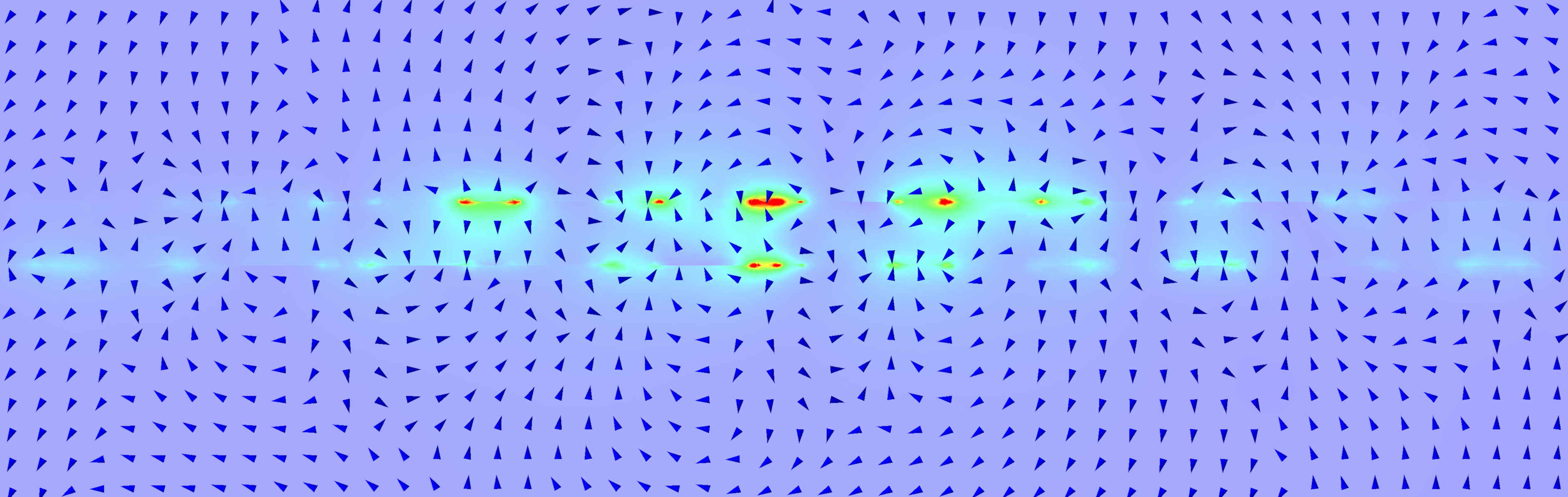}
			\label{Cut_2_S}	
				}
	\hspace{0.0005cm}
	\subfigure[]{
			\includegraphics[width=0.13\textwidth]{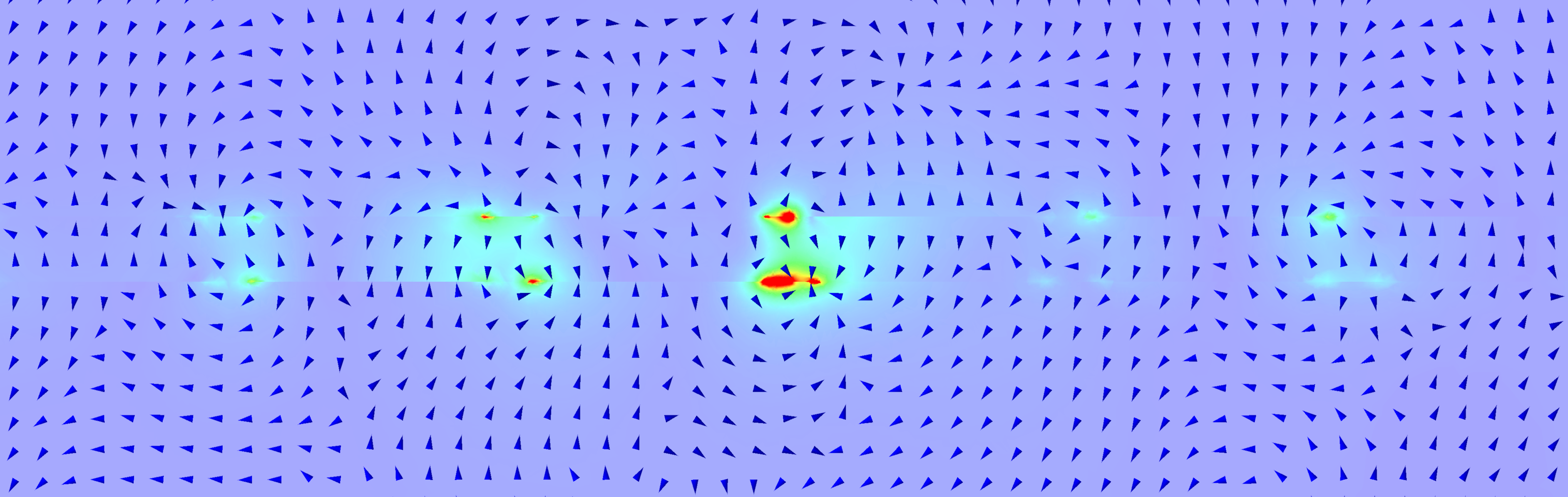}
			\label{Cut_3_S}	
	}
	\hspace{0.0005cm}
	\subfigure[]{
			\includegraphics[width=0.13\textwidth]{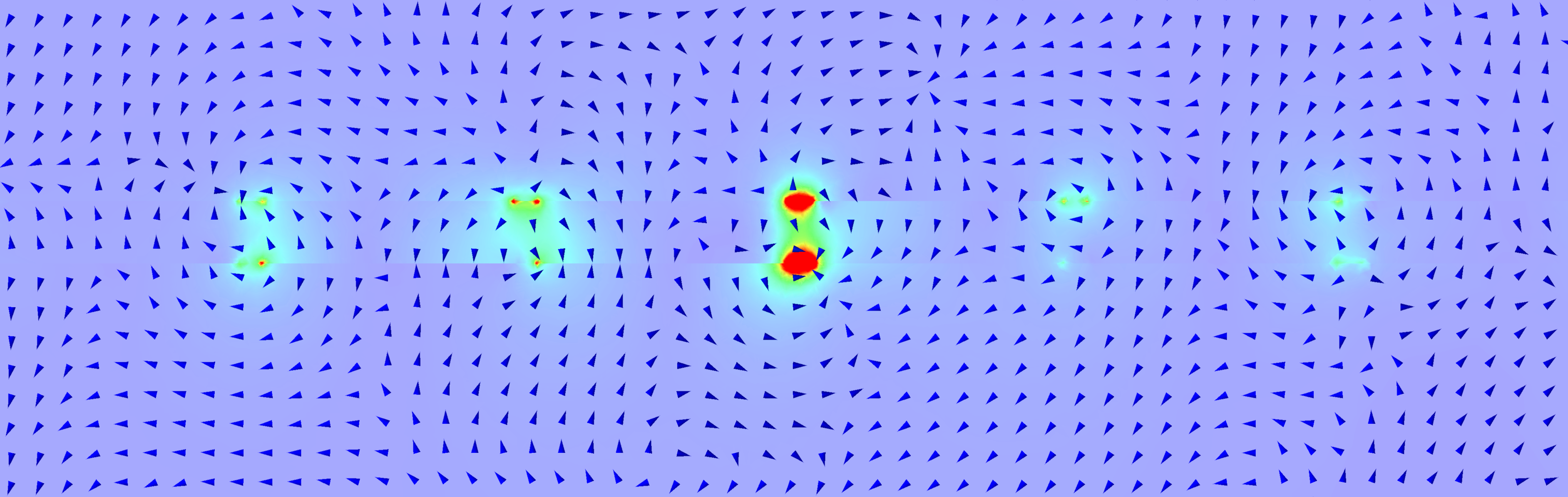}
			\label{Cut_4_S}	
				}
	\hspace{0.0005cm}
	\subfigure[]{
			\includegraphics[width=0.13\textwidth]{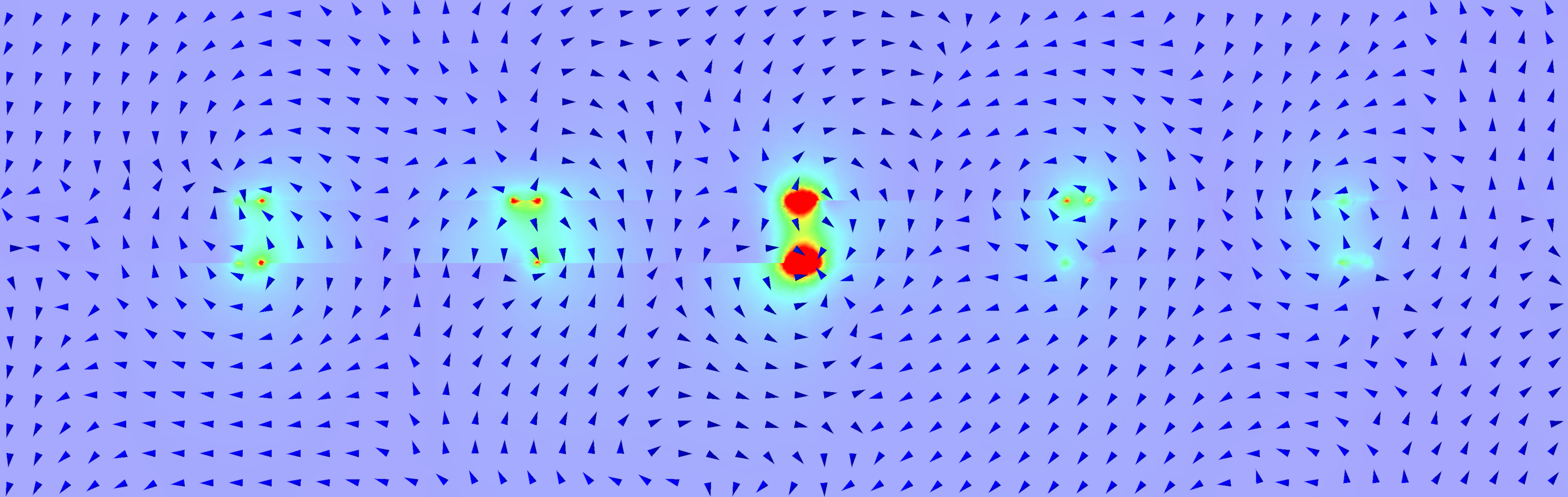}
			\label{Cut_5_S}	
	}
	\hspace{0.0005cm}
	\subfigure[]{
			\includegraphics[width=0.13\textwidth]{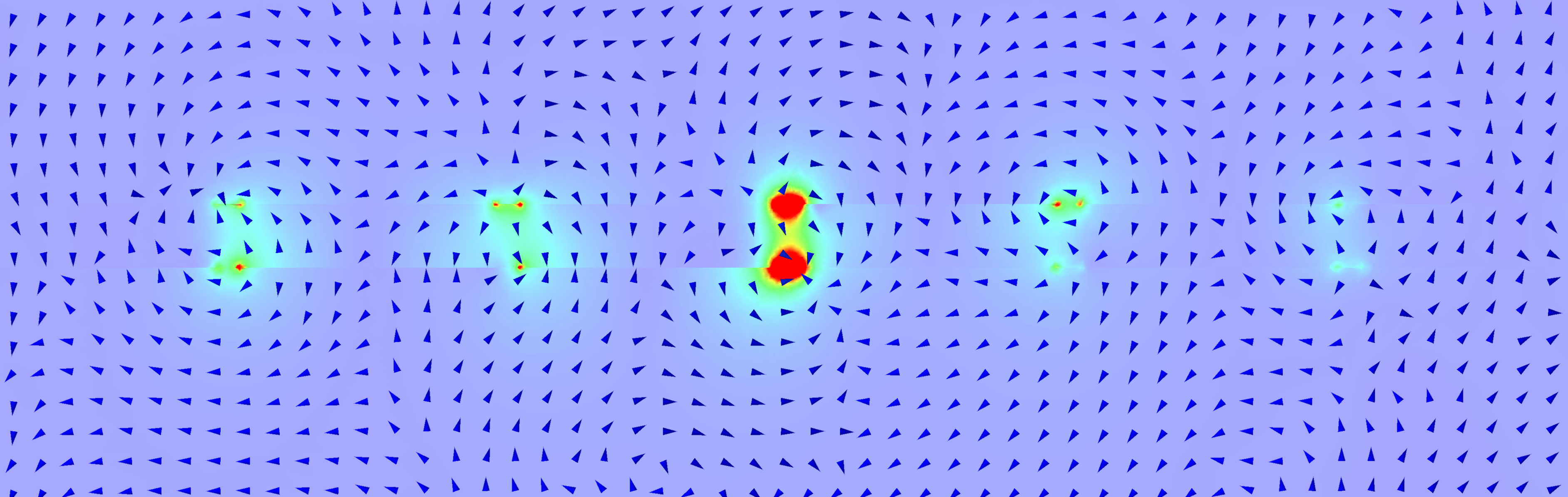}
			\label{Cut_6_S}	
				}
	\hspace{0.0005cm}
	\subfigure[]{
			\includegraphics[width=0.13\textwidth]{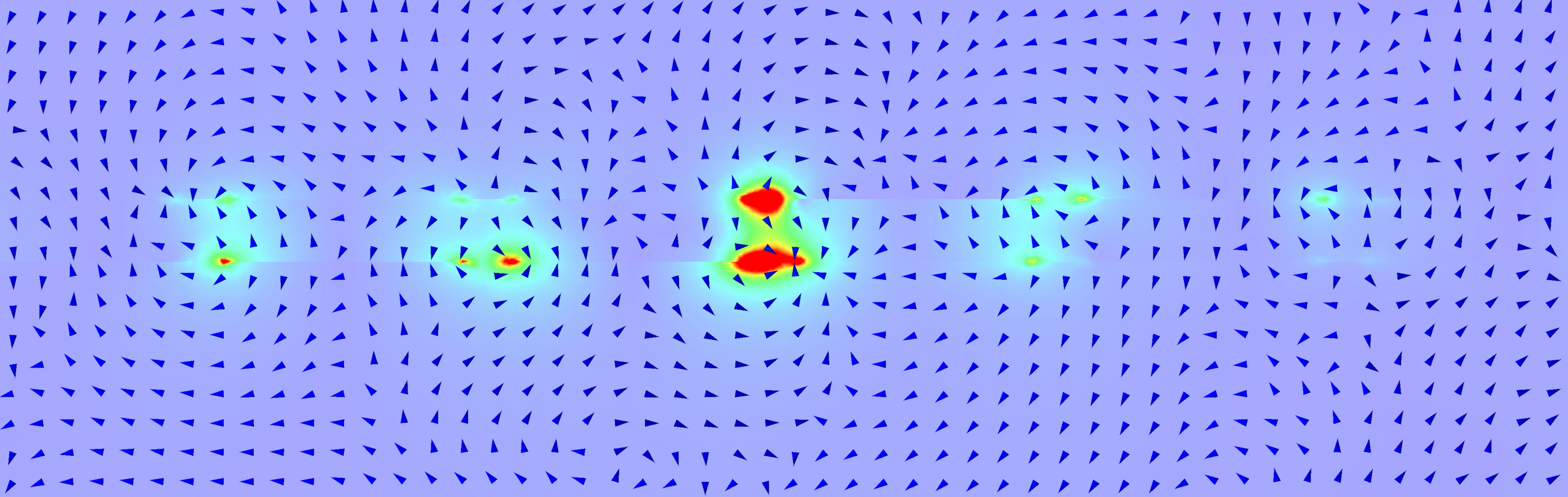}
			\label{Cut_7_S}	
	}
	\hspace{0.0005cm}
	\subfigure[]{
			\includegraphics[width=0.13\textwidth]{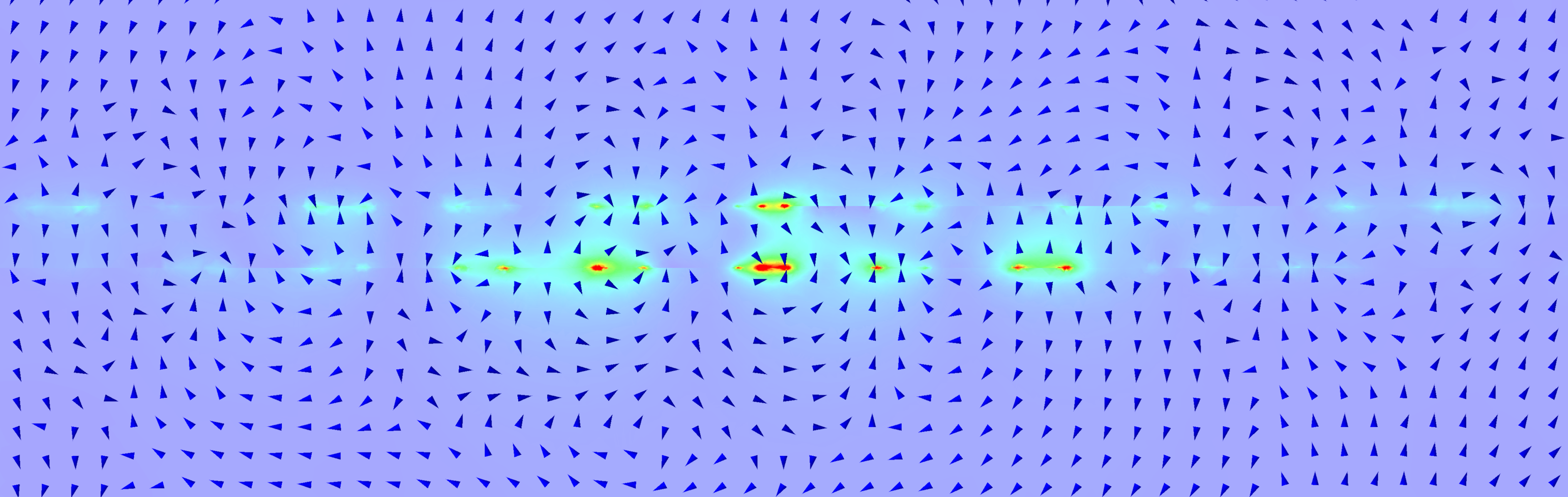}
			\label{Cut_8_S}	
				}
	\hspace{0.0005cm}
	\subfigure[]{
			\includegraphics[width=0.13\textwidth]{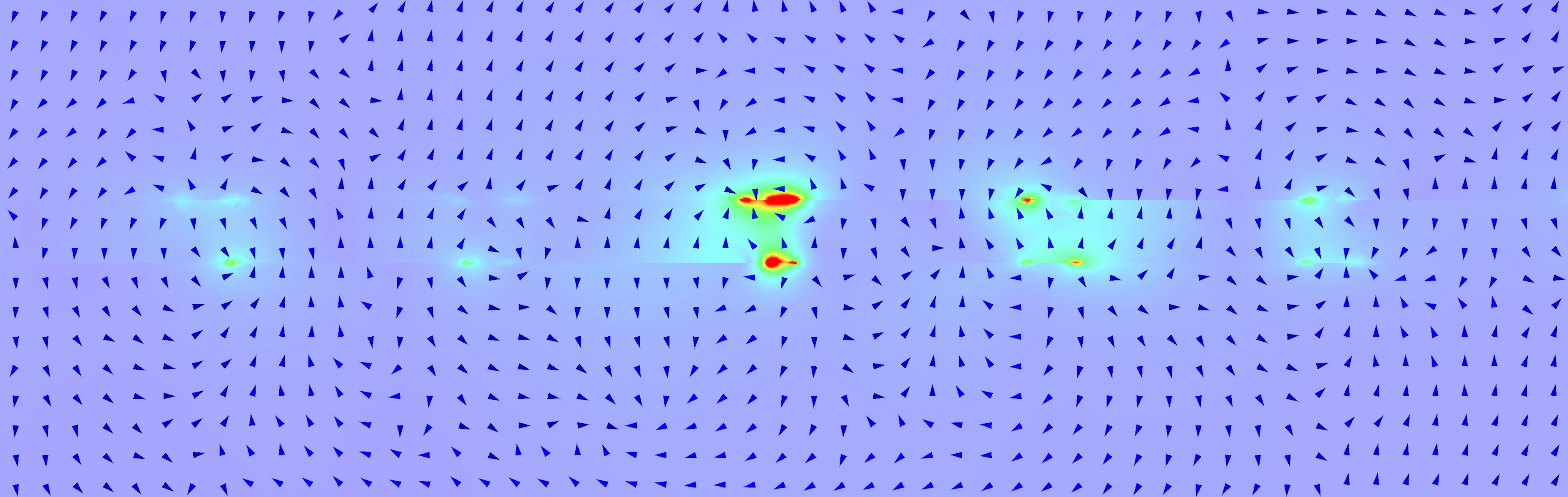}
			\label{Cut_9_S}	
	}
	\end{center}
	\caption{\subref{ASL_S} The field profile of an ASL line. 
	The field profiles of each cut line from \textcircled{1} to \textcircled{9} are depicted in \subref{Cut_1_S} to \subref{Cut_9_S}, as illustrated in Fig.\ref{raw_armchair_cutplane}.
	}
	\label{Cut_Planes}
\end{figure}

The field profile of the ASL and the identified cut planes are presented in Fig. \ref{Cut_Planes}, where the red and blue areas demonstrate the highest and the lowest amplitude of electric field,respectively. The direction of the triangles shows the direction of tangential electric field to the cut plane. All of these profiles exhibit confinement of the electric field in the interface of two side of structure but some of them have more similarity to the electric field of ASL cross section (see \ref{ASL_S}). Among these field profiles, those displayed in Figs. \ref{Cut_2_S}, \subref{Cut_8_S} show lesser similarity to the ASL profile in Fig. \ref{ASL_S} than the others, Because those cut lines are not perpendicular to the propagation direction.

The selection and cut the structure using these cut planes, would remains some fragmented parts which acting as local cavities, which disrupt the matching. 
Therefore, cut plane \textcircled{1} (or \textcircled{9}) has been selected despite having lesser similarity than cut planes of \textcircled{4} to \textcircled{7}, due to lesser effect of removing the fragmented parts. The dimensions of the ASL line play a decisive role in the coupling of two structures and would be calculated by optimization processes.
   
\begin{figure}[h!]
	\begin{center}
	\subfigure[]{\raisebox{5mm}{
			\includegraphics[width=0.45\textwidth]{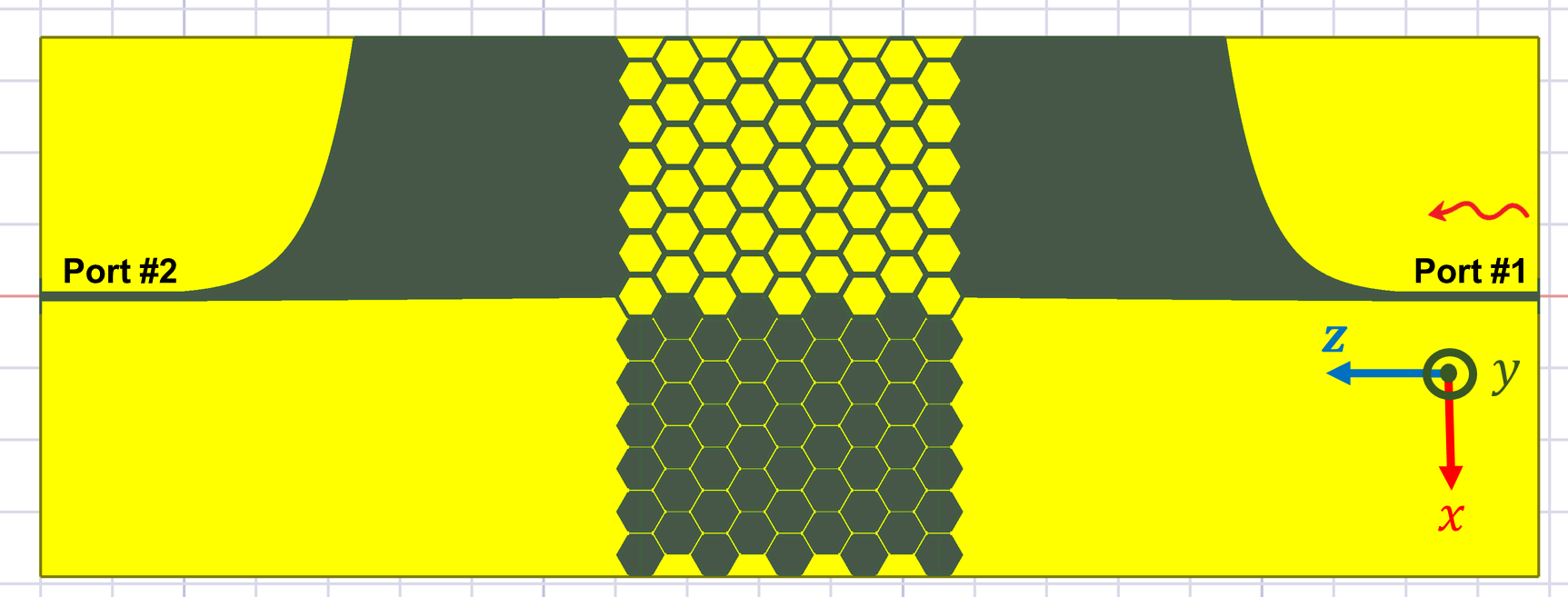}
			\label{transition_structure}	
				}
				}
	\hspace{0.0005cm}
	\subfigure[]{
			\includegraphics[width=0.48\textwidth]{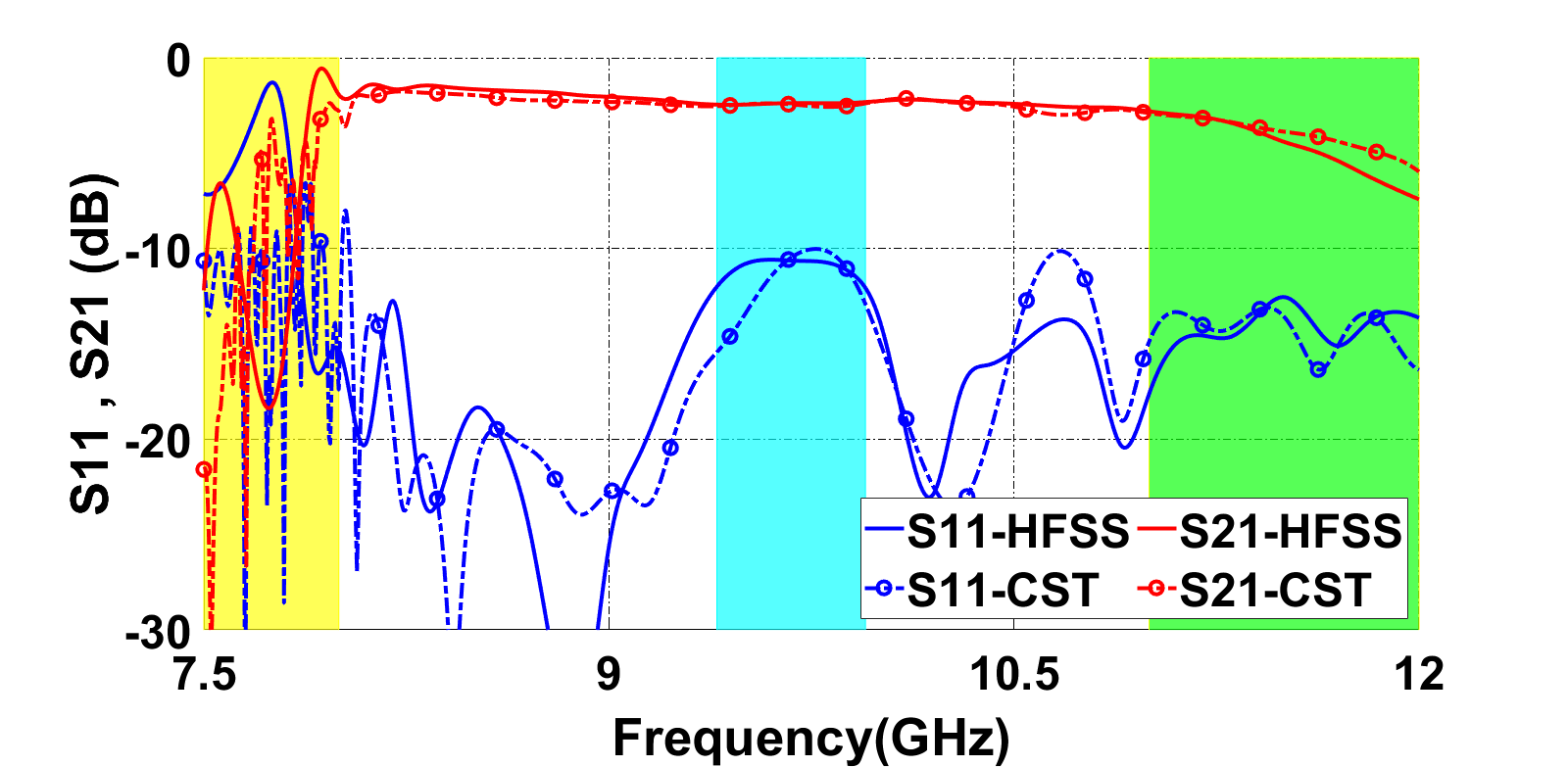}
			\label{Hex_armchair_transition}	
	}
	\end{center}
	\caption{\subref{transition_structure} The proposed structure utilizes cut plane \textcircled{1} from Fig.\ref{raw_armchair_cutplane}, with the remaining parts removed. \subref{Hex_armchair_transition} 
	The return and insertion loss of the proposed structure demonstrate an effective coupling between the ASL and the topological structure. The analysis has been conducted using both Ansys HFSS and CST Studio to validate the results, which align closely. In the yellow region, bulk modes are excited, causing the propagation to deviate from the line wave. The cyan region indicates the inherent bandgap of the armchair arrangement, where the return loss has improved. In the green region, there would be two edge simultaneously with opposite phase velocities.
	}
	\label{transition}
\end{figure}

The final structure and its scattering parameters are presented in Fig. \ref{transition}, demonstrating an effective coupling between the ASL and the topological structure. This finding aligns closely with the dispersion diagram of the ribbon shown in Fig \ref{Armchair_Ribbon_dispersion_a_12_b_05}.
The simulation results obtained using Ansys HFSS were validated by conducting repeated simulations with CST Studio, demonstrating that the results are in good agreement.

In Fig. \ref{Hex_armchair_transition}, the bulk mode regions are highlighted in yellow, indicating high return loss as expected. The cyan region represents the range in which a complete bandgap exists for the armchair configuration. Despite the presence of the bandgap, the insertion loss exhibits only slight variations in this range, while the return loss increases. In the green region, the insertion loss decreases due to the simultaneous presence of two edge modes in opposite phase velocities.
 
\begin{figure}[h!]
	\begin{center}
			\includegraphics[width=0.4\textwidth]{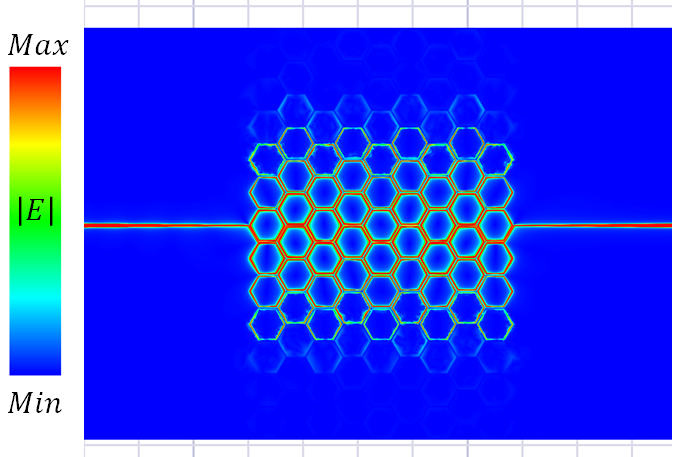}
	\end{center}
	\caption{ 
	The electric field profile propagating through the structure is shown in at 7.9 GHz, where it is located in the bulk mode region and permeates through the bulk rather than concentrated at the interface.
	}
	\label{bulk_modes}
\end{figure}

The electric field profile in the proposed structure is shown for the frequency of 7.9 GHz in Fig. \ref{bulk_modes}. The frequency correspond to the bulk modes, where the electric field permeates through the bulk. This result provides further validation for the previously mentioned calculations of the ribbon dispersion diagram and the scattering parameters.
  
\section{Antenna application}
The main advantage of using the armchair arrangement in antenna applications is that it effectively places most of the topological bandgap within the fast wave region. According to the findings in Appendix A, this design enables the proposed structure to scan through the broadside, a capability that conventional unbalanced CRLH leaky-wave antennas do not offer.



\begin{figure}[h!]
\begin{center}
\subfigure[]{
 \includegraphics[width=0.45\textwidth]{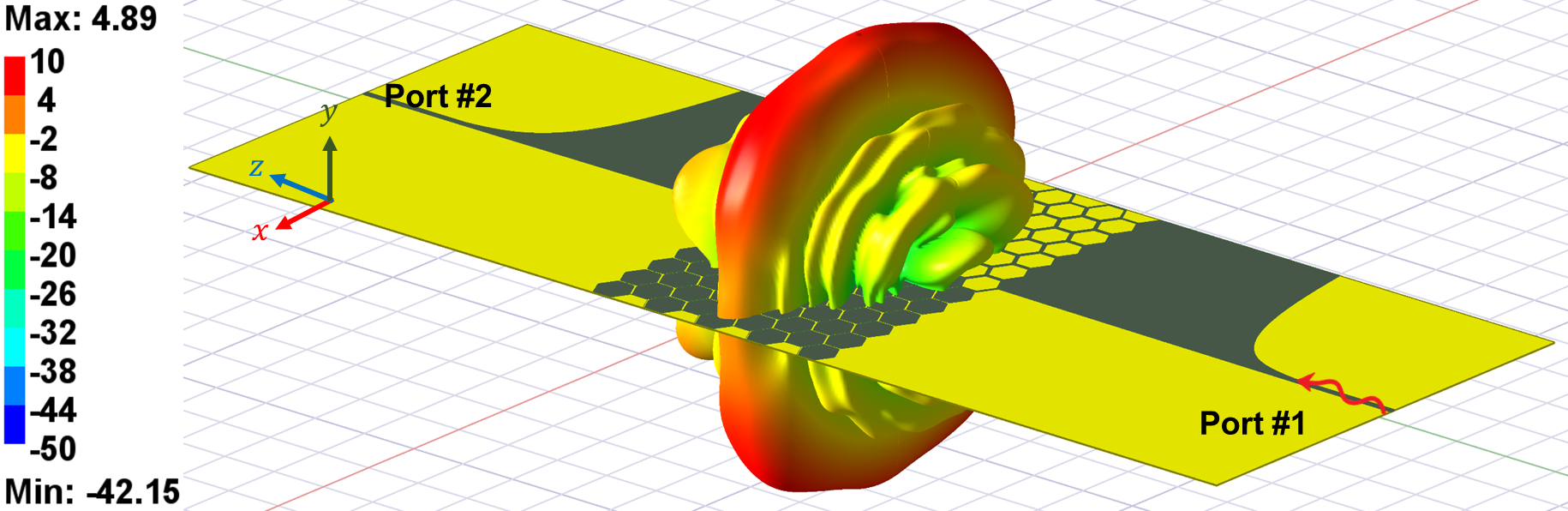}
\label{Antenna_wholeScoorS}
}
\hspace{0.0005cm}
\subfigure[]{
\includegraphics[width=0.45\textwidth]{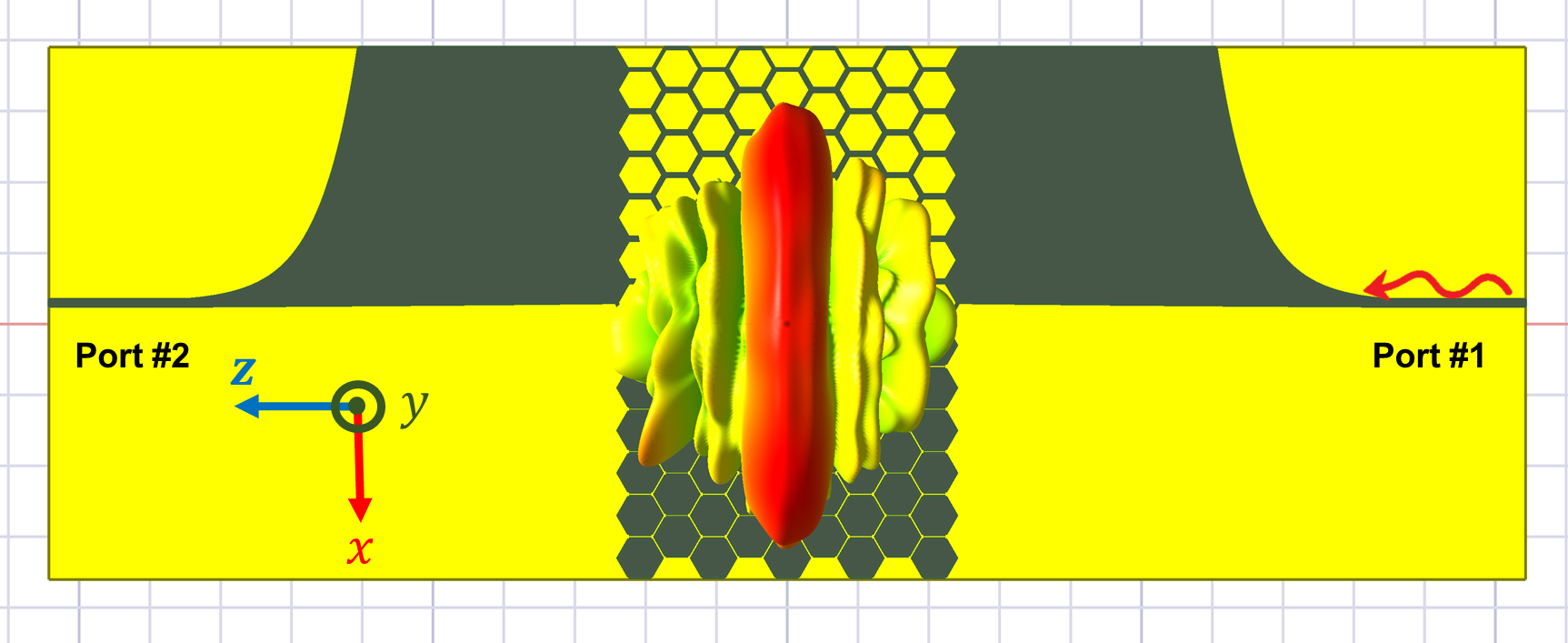}
\label{Antenna_whole_from_aboveScoorS}
}
\end{center}
\caption{\subref{Antenna_wholeScoorS}The realized gain pattern of the proposed antenna at 9.5 GHz radiates in the broadside direction.\subref{Antenna_whole_from_aboveScoorS} A top view of the pattern is provided to better illustrate the direction of radiation. }
\label{Pattern3D}
\end{figure}

\begin{figure}[h!]
	\begin{center}
			\includegraphics[width=0.4\textwidth]{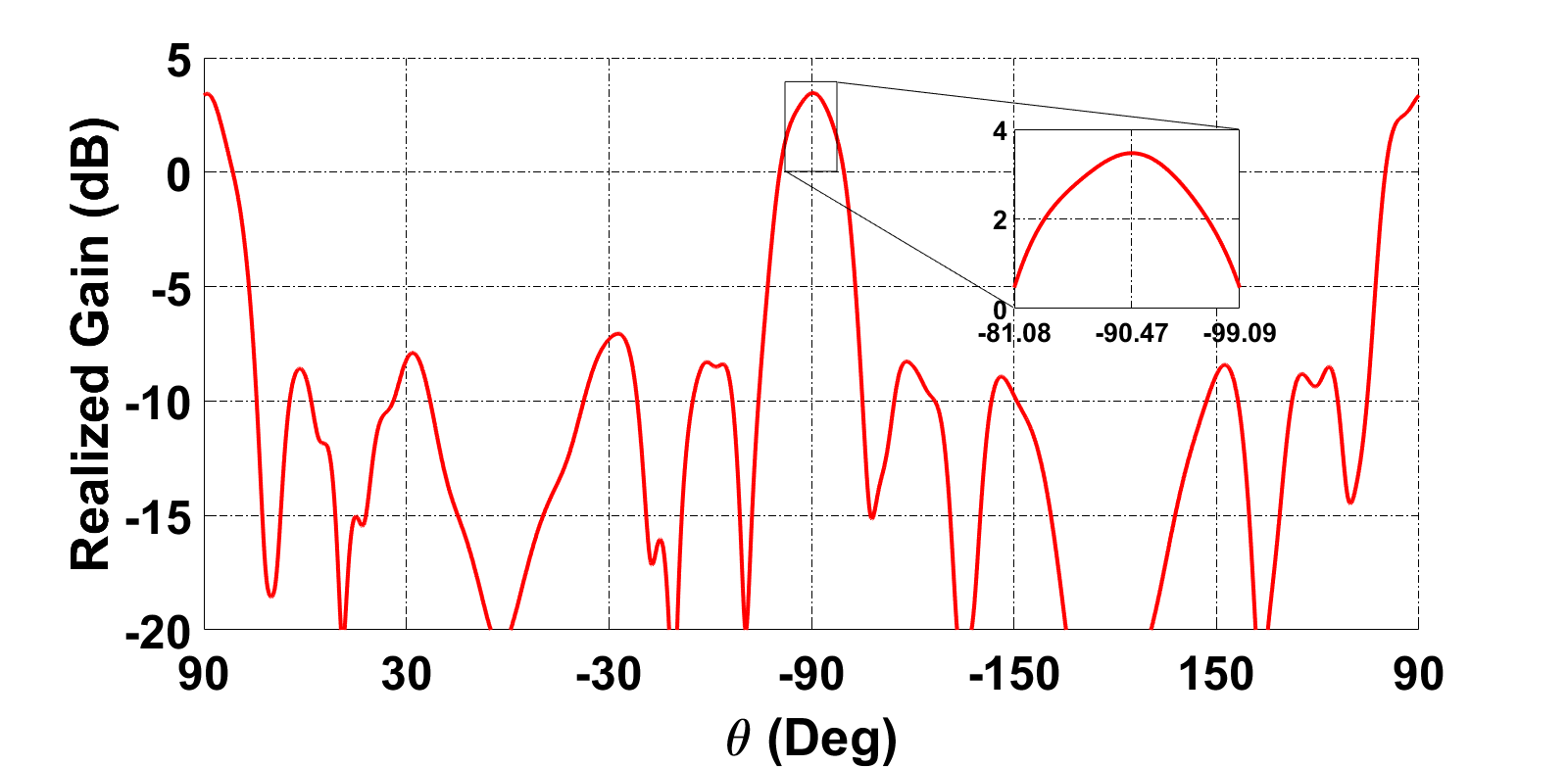}
	\end{center}
	\caption{The realized gain of the proposed structure in the $\phi=90^{\circ}$ plane at 9.5 GHz is approximately 3.47 dB, with a HPBW of around 18 degrees, as shown in the inset plot.   
	}
	\label{rect9.5}
\end{figure}

The 3D pattern of the proposed antenna operating at 9.5 GHz is shown in Fig. \ref{Pattern3D}, along with its structure. The design demonstrates equal radiation from the broadside of both sides. The length of the radiating portion is approximately 97 mm; however, extending this length could lead to higher gains.
To enhance understanding, the 2D pattern shown in Fig.\ref{Pattern3D} for the $\phi=90^{\circ}$ plane is represented in Fig.\ref{rect9.5}.
The radiation is directed completely broadside, with a realized gain of 3.47 dB, a HPBW of 18 degrees, and a SLL of less than -10 dB.

\begin{figure}[h!]
	\begin{center}
	\subfigure[]{
			\includegraphics[width=0.2\textwidth]{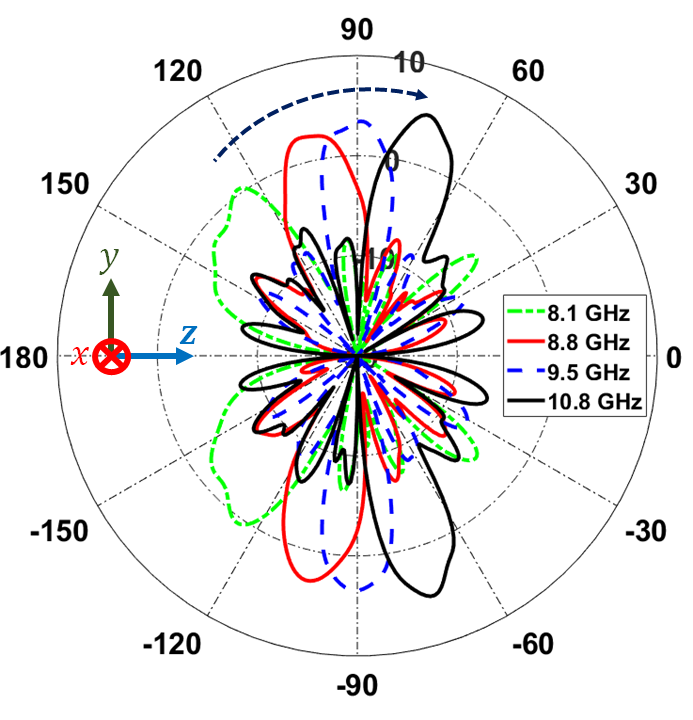}
			\label{RealizedGain_Polar_Total_flesh_HFSS}	
				}
	\hspace{0.0005cm}
	\subfigure[]{
			\includegraphics[width=0.2\textwidth]{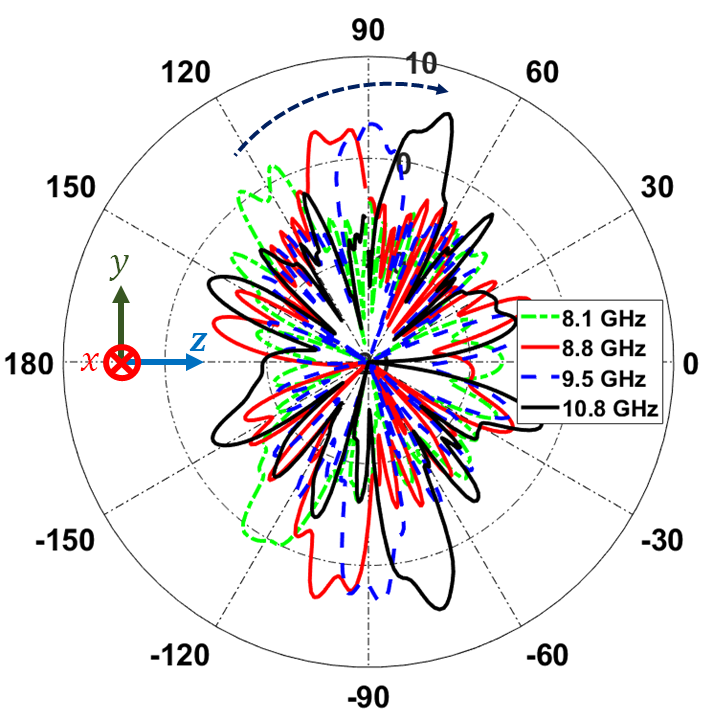}
			\label{RealizedGain_Polar_Total_flesh_CST}	
	}
	\hfill
	\subfigure[]{
		\includegraphics[width=0.2\textwidth]{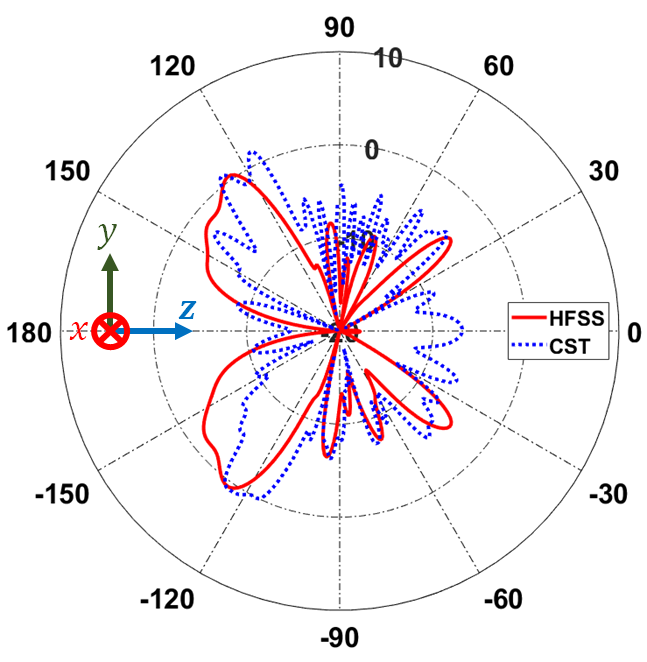}
		\label{8.1GHz}
		}
		\hfill
	\subfigure[]{
		\includegraphics[width=0.2\textwidth]{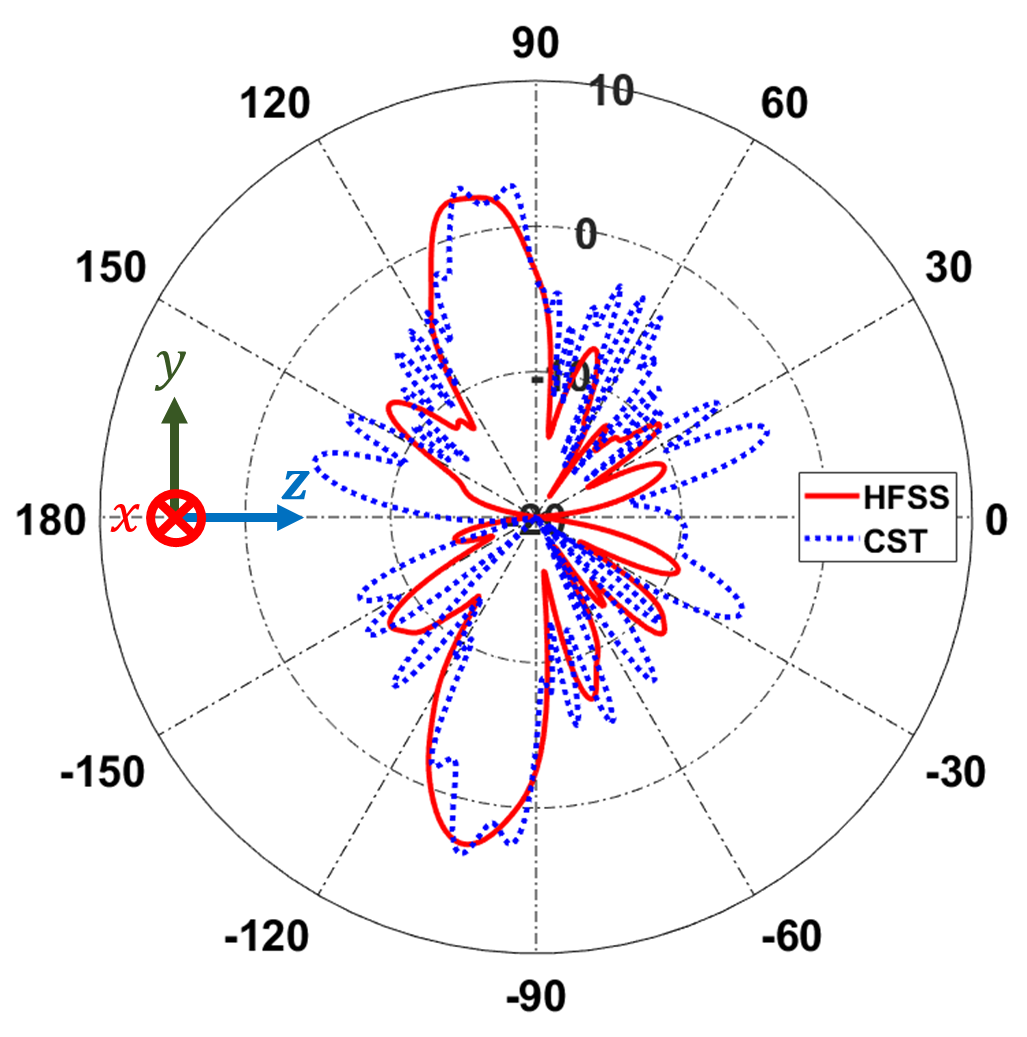}
		\label{8.8GHz}
		}
		\hfill		
	\subfigure[]{
		\includegraphics[width=0.2\textwidth]{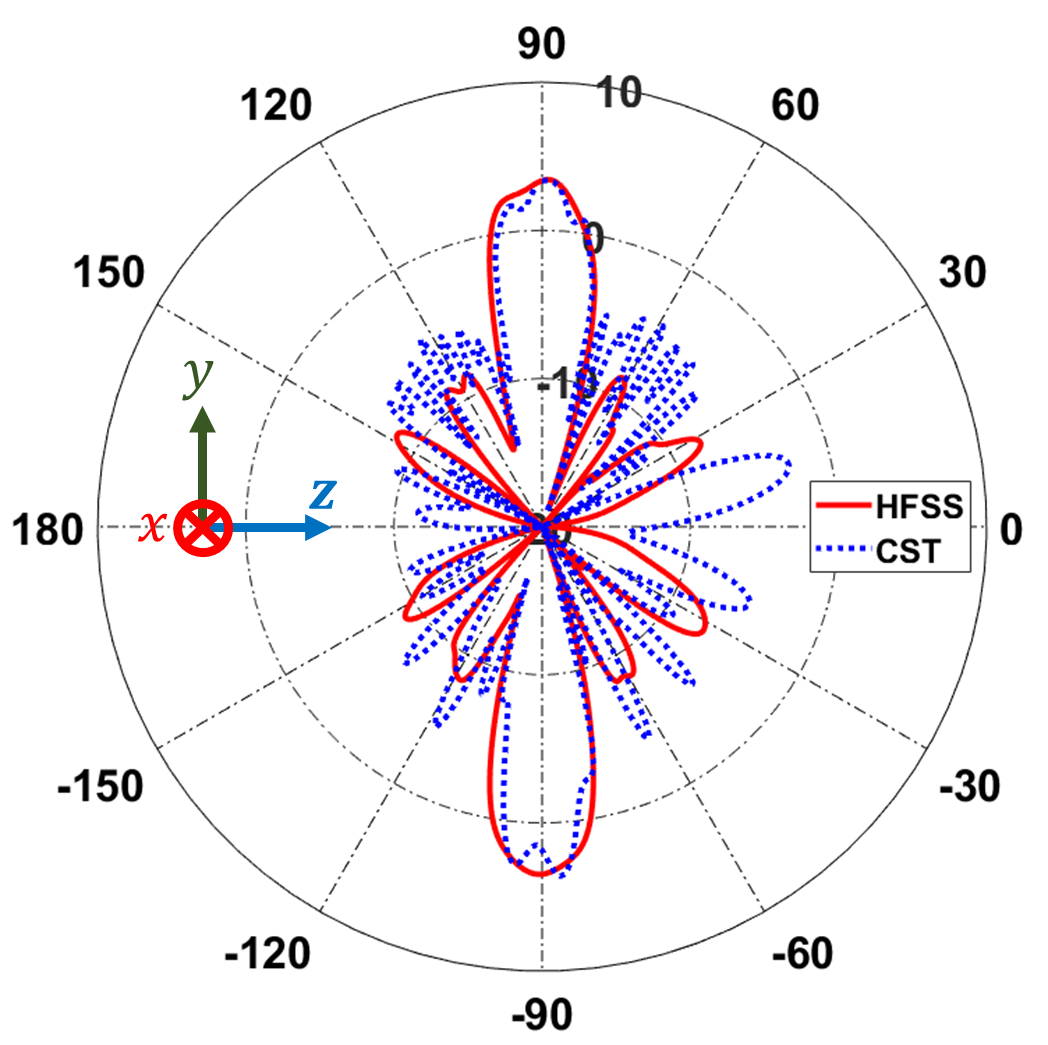}
		\label{9.5GHz}
		}
		\hfill
	\subfigure[]{
		\includegraphics[width=0.2\textwidth]{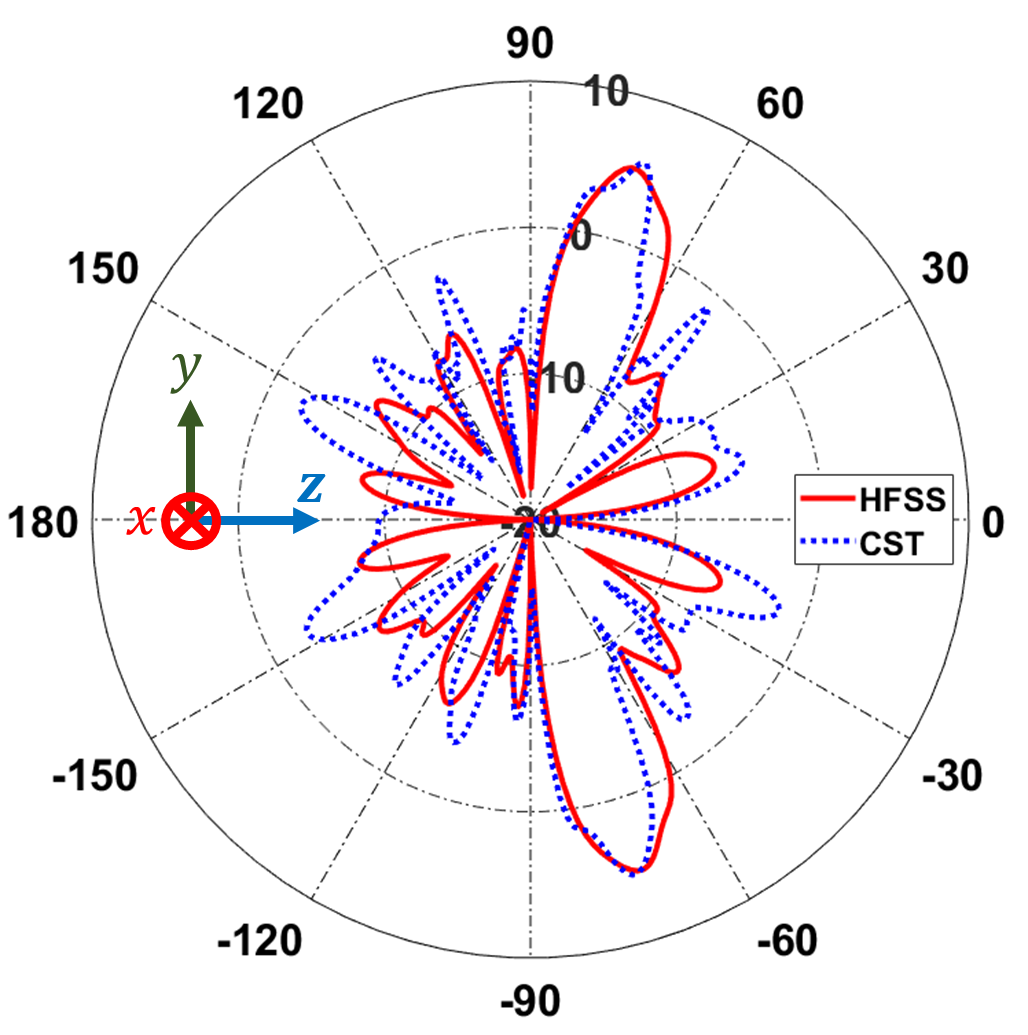}
		\label{10.8GHz}
		}
	\end{center}
	\caption{ 
	The realized gain pattern of the proposed antenna in the $\phi=90^{\circ}$ plane for the frequencies of 8.1, 8.8, 9.5 and 10.8 is shown using \subref{RealizedGain_Polar_Total_flesh_HFSS} Ansys HFSS and \subref{RealizedGain_Polar_Total_flesh_CST} CST studio. The arrow shows the direction of the main lobe scan as the frequency increases. \subref{8.1GHz} to \subref{10.8GHz} are included for improved clarity, showcasing the correlation between the results of the two commercial software packages. The slight discrepancies observed in \subref{8.1GHz} are attributed to the limited excitation of bulk modes and energy transference from the interface. 
	}
	\label{RealizedGain_Polar_Total}
\end{figure}

The antenna features a dual beam pattern on both sides, scanning the space in a similar manner. The backward radiation starts at approximately 8 GHz with an angle of about $\pm 126$ degrees. As the frequency increases, the pattern crosses the broadside and reaches the end of its scanning range at an angle of approximately $\pm 73$ degrees, around 10.8 GHz. Fig.\ref{RealizedGain_Polar_Total} illustrates the realized gain patterns obtained from Ansys HFSS and CST Studio.

\begin{figure}[h!]
	\begin{center}
			\includegraphics[width=0.4\textwidth]{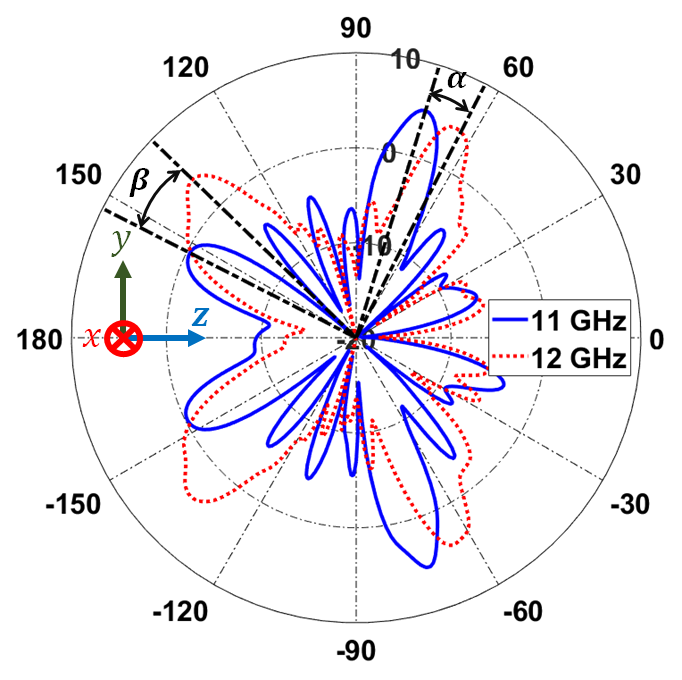}
	\end{center}
	\caption{ 
	The realized gains of the proposed structure in the $\phi = 90^{\circ}$ plane at 11 and 12 GHz show two additional beams directed backward. The scanning rates differ between the forward and backward beams; specifically, with each 1 GHz increase in frequency, the angle \(\alpha\) for the forward beam is approximately \(10^{\circ}\), while the angle \(\beta\) for the backward beams is approximately \(17^{\circ}\). These results align with the dispersion diagram of \ref{Selected_Ribbon}, which indicates two modes with opposite phase velocities.
	}
	\label{Polar_HFSS_11___12_dir}
\end{figure}

As the frequency exceeds approximately 11 GHz, two additional beams will emerge in the backward direction due to the presence of two edge modes with opposite phase velocities, as shown in Fig. \ref{Selected_Ribbon}.  The realized gain patterns at frequencies of 11 GHz and 12 GHz are illustrated in Fig. \ref{Polar_HFSS_11___12_dir}, showing the scanning of both backward and forward beams at different rates.

To prevent beams from being directed simultaneously in backward and forward directions, the antenna's operating bandwidth should be limited to approximately 8 to 11 GHz, where the light line intersects with the edge mode. 

\begin{figure}[h!]
	\begin{center}
			\includegraphics[width=0.4\textwidth]{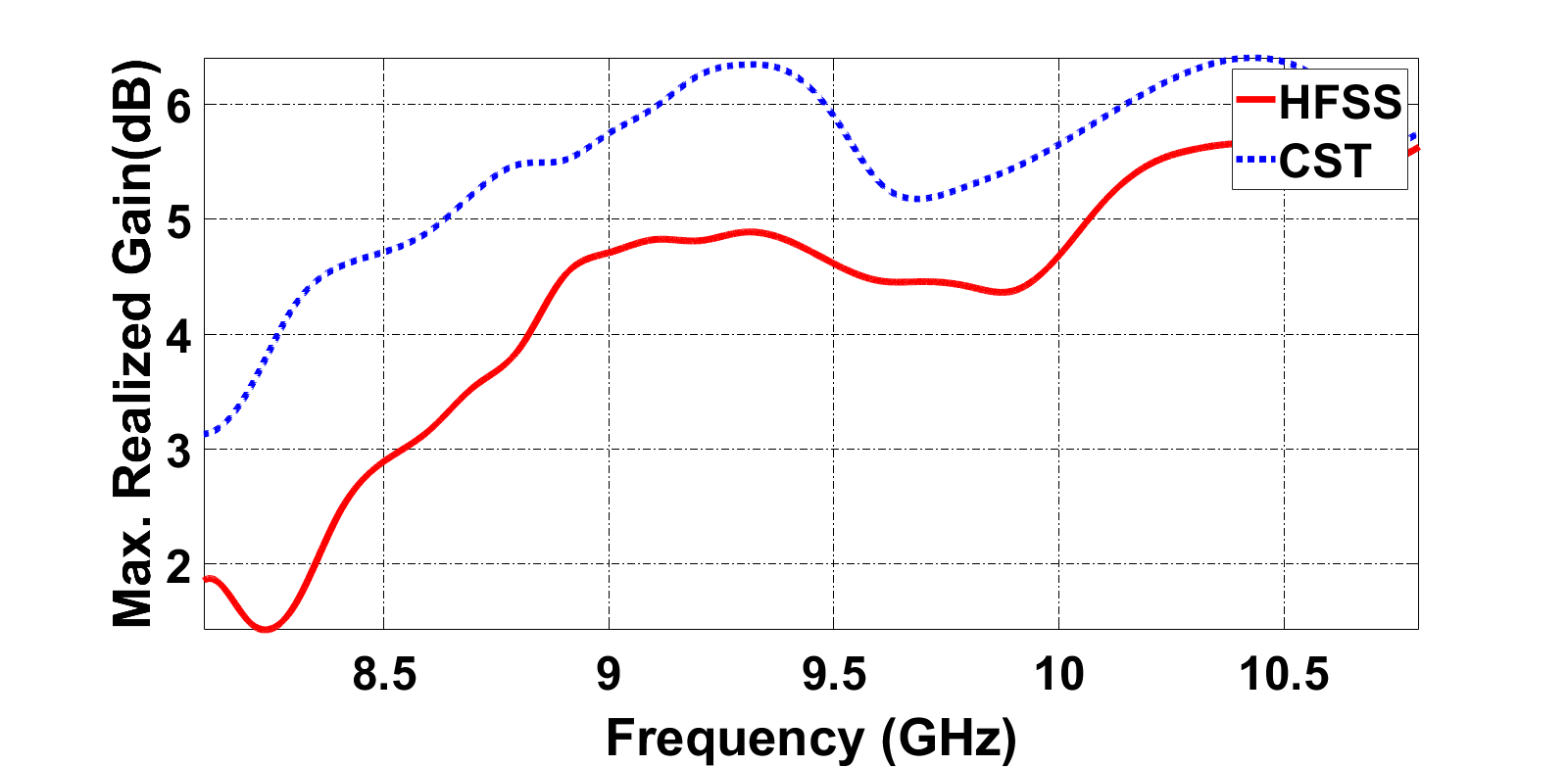}
	\end{center}
	\caption{The maximum realized gain of the structure is observed for both Ansys HFSS and CST Studio as the frequency increases. The results are consistent, confirming their accuracy. In most of the frequency range, the maximum realized gain shows only slight variations. The minor decrease observed in the middle of this range occurs due to a gap in the dispersion diagram of the ribbon, as illustrated in Fig.\ref{Selected_Ribbon}.     
	}
	\label{RealizedGain_vs_freq_HFSS_and_CST}
\end{figure}

The variation of maximum realized gain with increasing frequency, as analyzed using Ansys HFSS and CST Studio, is illustrated in Fig. \ref{RealizedGain_vs_freq_HFSS_and_CST}. The results are consistent and provide further validation of the simulations.
The slight decline observed in the frequency range of 9.5 to 10 GHz is attributed to the inherent bandgap of the armchair arrangement. In contrast, to the best of our knowledge, other unbalanced CRLH leaky-wave antennas do not exhibit this characteristic \cite{monticone, caloz2005, leaky, jackson_leaky}. Another noteworthy feature is the nearly constant realized gain across most of the operational frequency range, which typically ranges from 4 to 6 dB. 

\section{Conclusions}\label{sec5}


A leaky-wave antenna is proposed using spin photonic topological insulators (PTIs). It features a hexagonal complementary unit cell arranged in an armchair configuration. The primary advantage of using the armchair arrangement in leaky-wave antenna applications is that most of topological bandgap is positioned within the fast wave region. In contrast, for the zigzag arrangement, most of the topological bandgap falls within the slow wave region. The design of the proposed antenna allows for a scanning range of 53 degrees, from $\pm$126 to $\pm$73 degrees, within an approximate bandwidth of 2.7 GHz. Importantly, the antenna sustains its performance at broadside angles without experiencing a significant decrease, even in the presence of an open stopband. 


In Table \ref{tableC}, a comparison of the proposed leaky-wave antennas utilizing topological insulators is presented. The table indicates that the antenna designed with an armchair arrangement offers the widest bandwidth and the highest scanning range.
The HPMW of the proposed antenna is a weaker point compared to other references, particularly \cite{abtahi2} and \cite{chernAntenna}. This limitation can be improved by increasing the antenna's length. None of the references listed in the table can radiate in a broadside direction, except for \cite{chernAntenna}. However, this option has significant drawbacks, including a single beam pattern, a narrow bandwidth, a bulky design, and the requirement for a static magnetic field bias. In contrast, the proposed antenna not only radiates broadside but also offers two beams, a wide bandwidth, a low-profile design, and does not require any additional facilities. 

\renewcommand{\arraystretch}{1.6}
\begin{table*}[!h]
	\begin{center}
		\caption{comparison between leaky-wave antennas based on PTI.}
		\begin{tabular}{ccccccc}
				 				& Type & Lattice     & Bandwidth       & HPBW$^1$         & Scanning Range$^2$     & Length   \\ \hline 
			\cite{chernAntenna} & Chern PTI & square & 6.34-6.76 GHz   & $4.5^{\circ}$ & $\approx 70^{\circ}$ to $110^{\circ}$& 525 mm    \\ \hline
			
			\cite{chernantennaPeriodic} & Chern PTI & square & 13.2-14 GHz   & $10^{\circ}$ & $\approx 52^{\circ}$ to $40^{\circ}$& 156 mm    \\ \hline
			
			\multirow{ 2}{*}{\cite{SinghAntenna} } & \multirow{ 2}{*}{Spin PTI} & Hexagonal  & \multirow{ 2}{*}{19.75-21.25 GHz} & \multirow{ 2}{*}{$\approx 6^{\circ}$}& $+166^{\circ}$ to $+130^{\circ}$  & \multirow{ 2}{*}{	$\approx$ 168 mm}  \\ 
								&		& Zig-Zag	&				  &				  &	$-166^{\circ}$ to $-130^{\circ}$  &            \\ \hline
								
			\multirow{ 2}{*}{\cite{abtahi2} }& \multirow{ 2}{*}{Spin PTI} & 30 degree & \multirow{ 2}{*}{17.5-20 GHz} & \multirow{ 2}{*}{$5^{\circ}$} & $+122^{\circ}$ to $+99.5^{\circ}$& \multirow{ 2}{*}{200 mm}   \\ 
								&			& Rhombic & 	 &				  & $-122^{\circ}$ to $-99.5^{\circ}$& 	    \\ \hline
								
			\multirow{ 2}{*}{This work }& \multirow{ 2}{*}{Spin PTI} & Hexagonal & \multirow{ 2}{*}{8.1-10.8 GHz} & \multirow{ 2}{*}{$18^{\circ}$} & $+126^{\circ}$ to $+73^{\circ}$& \multirow{ 2}{*}{97 mm}   \\ 
								&			&  Armchair & 	  &				  & $-126^{\circ}$ to $-73^{\circ}$& 	    \\ \hline
			\multicolumn{7}{l}{$^1$ Minimum value in scan range is reported}\\
			\multicolumn{7}{l}{$^2$ The broadside is at $90^{\circ}$}\\
			
		\end{tabular}
		\label{tableC}
	\end{center}
\end{table*}

\appendices
\section{\break The Zig-Zag and Armchair Arrangements Integration}
\label{app1}
\begin{figure}[h!]
\begin{center}
\subfigure[]{
 \includegraphics[width=0.45\textwidth]{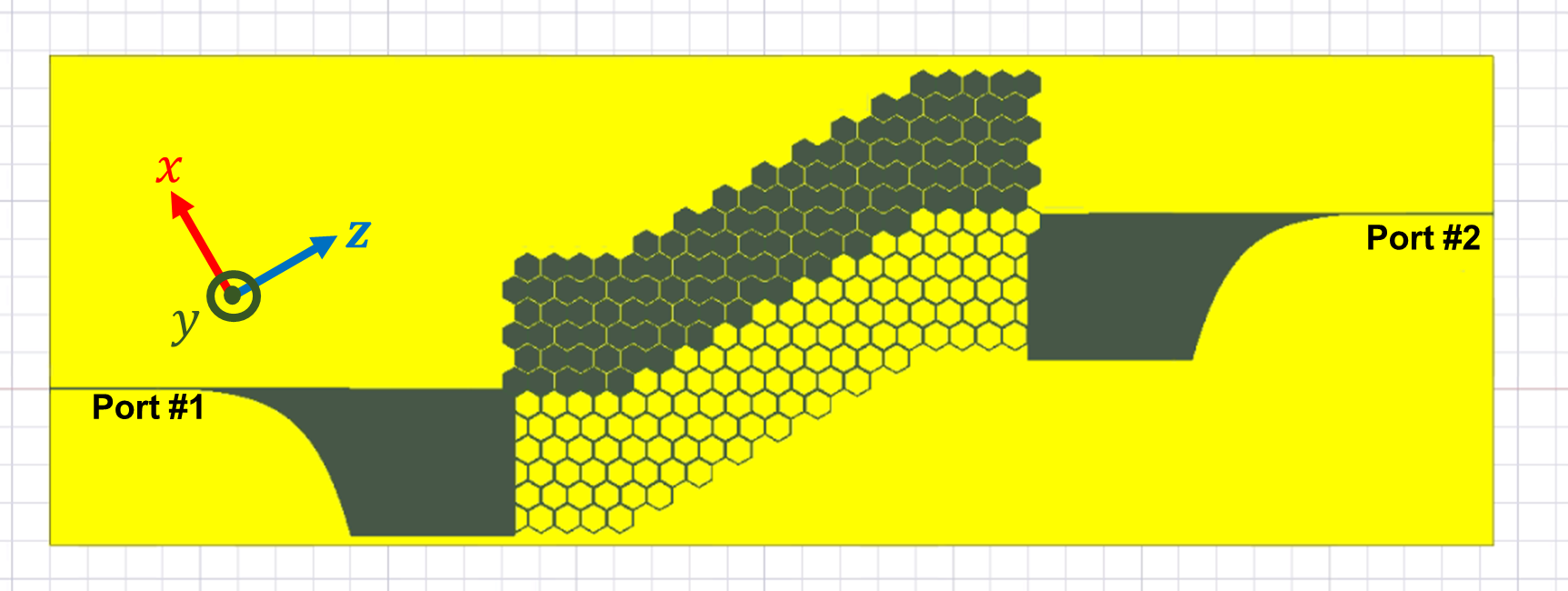}
\label{Antstruc}
}
\hspace{0.0005cm}
\subfigure[]{
\includegraphics[width=0.45\textwidth]{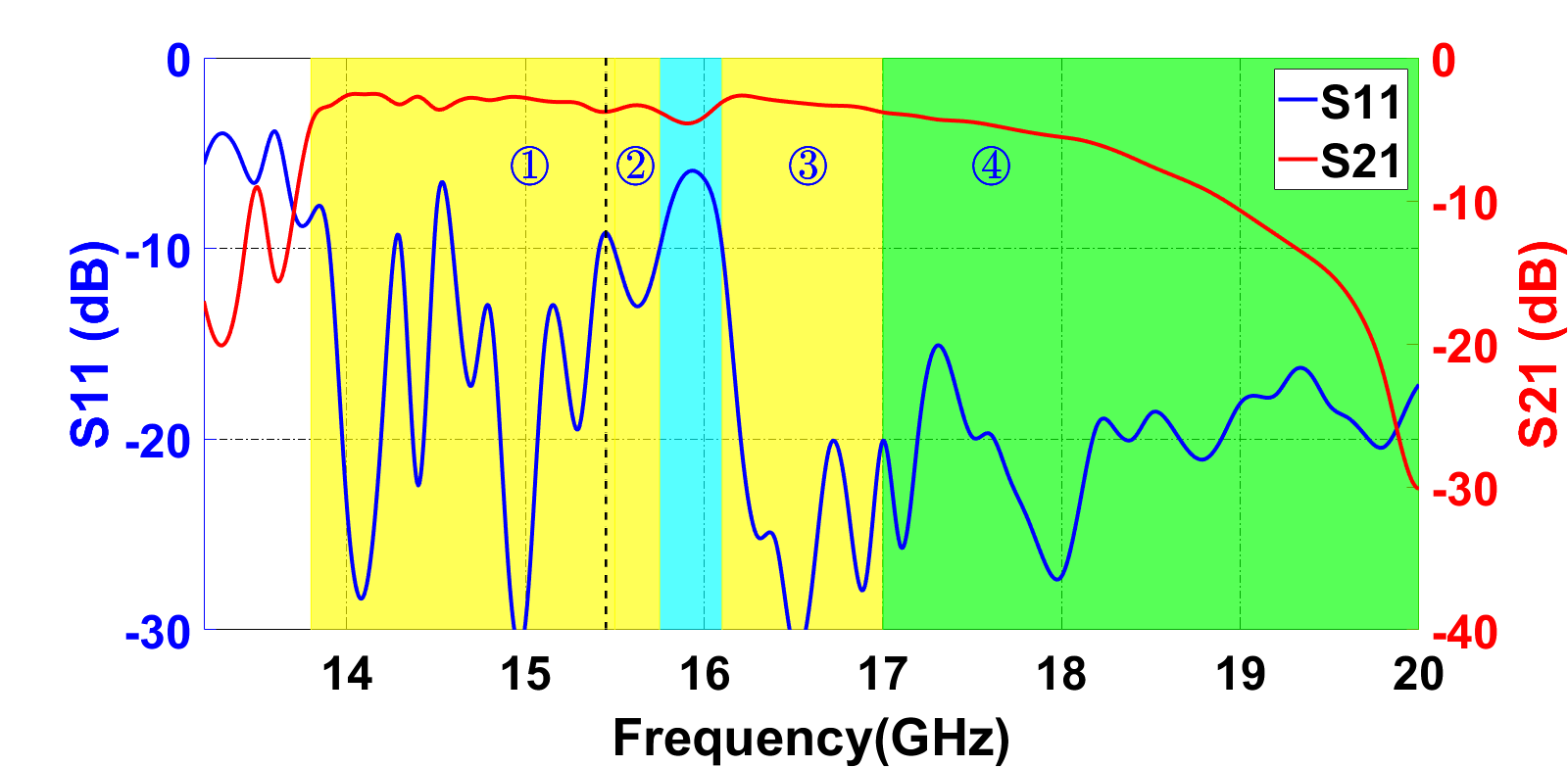}
\label{S_Para}
}
\end{center}
\caption{\subref{Antstruc}The antenna configuration by sandwiching the armchair arrangement between two zigzag counterparts and excitation with ASL.\subref{S_Para}The Scattering parameters of the proposed antenna.The yellow areas represent the frequencies at which only the armchair portion of the antenna structure radiates. In the green area, both the armchair and zigzag parts radiate simultaneously. The cyan regions indicate the gap in the edge mode, where the return loss increases. }
\label{Appendix_Pattern3D}
\end{figure}

The antipodal slot line (ASL) excitation method was introduced to integrate classical line structures with topological configurations in \cite{davisASL}. However, this method may cause coupling between the ASL and topological structures arranged in a zigzag configuration. As a preliminary step to analyze the behavior of the structure in an armchair arrangement, it was placed between two peripheral zigzag structures (see Fig.\ref{Antstruc}).

Fig.\ref{S_Para} illustrates the antenna's return and insertion loss, which corresponds with the dispersion diagrams shown in  Fig.\ref{Ribbon}. The bandwidth of exclusive radiation from the armchair section is highlighted in yellow.While in the green frequency interval, both zigzag and armchair configurations emit radiation. The cyan regions represent the gap in the edge mode, where the return loss increases.

The yellow and green regions are divided into four intervals, and the behavior of the structure changes in each with respect to whether it is right-handed or left-handed.

In domain number $\raisebox{.5pt}{\textcircled{\raisebox{-.9pt} {1}}}$, the armchair section is left-handed, while the zigzag section is right-handed. In domain number $\raisebox{.5pt}{\textcircled{\raisebox{-.9pt} {2}}}$, both the armchair and zigzag sections are left-handed. In domain number $\raisebox{.5pt}{\textcircled{\raisebox{-.9pt} {3}}}$, the armchair section is right-handed, and the zigzag section is left-handed. In domain number $\raisebox{.5pt}{\textcircled{\raisebox{-.9pt} {4}}}$, the behavior is similar to that of domain $\raisebox{.5pt}{\textcircled{\raisebox{-.9pt} {3}}}$, but the zigzag section operates in the fast wave regime.

\begin{figure}[h!]
	\begin{center}
	\subfigure[]{
	\includegraphics[width=0.4\textwidth]{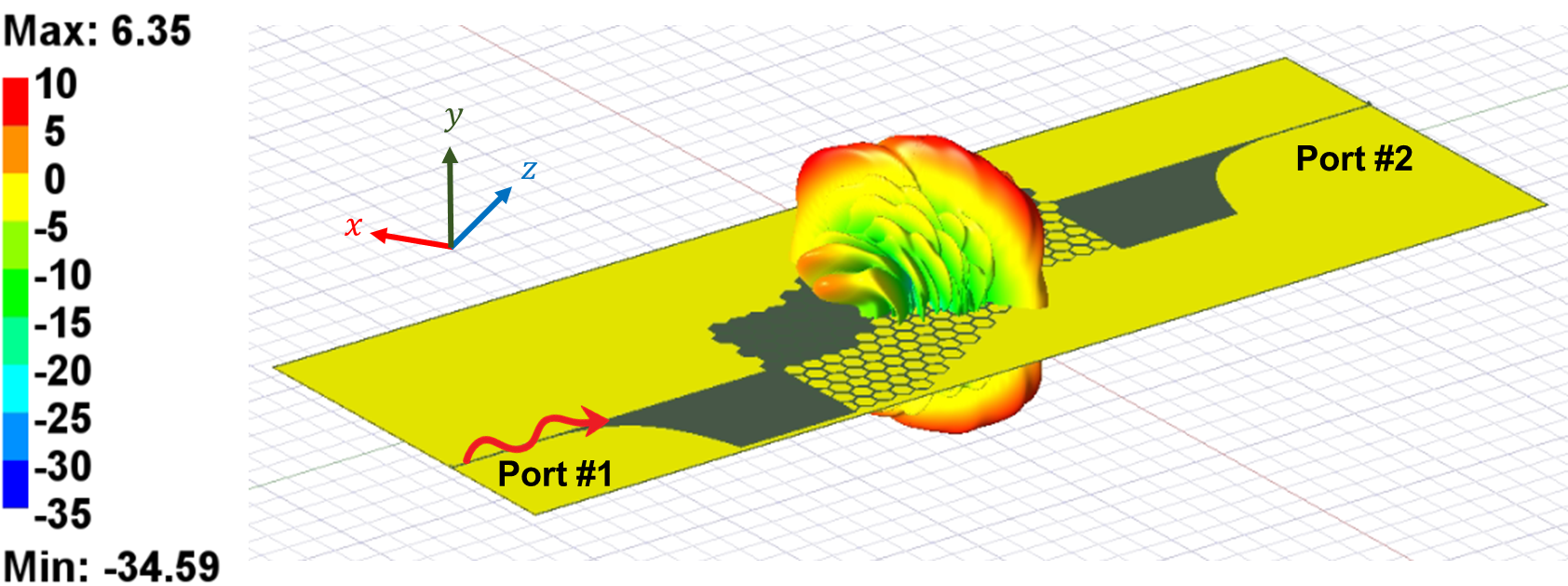}
	\label{ArmAntPatPort1}
	}
	\hspace{0.0005cm}
	\subfigure[]{
	\includegraphics[width=0.3\textwidth]{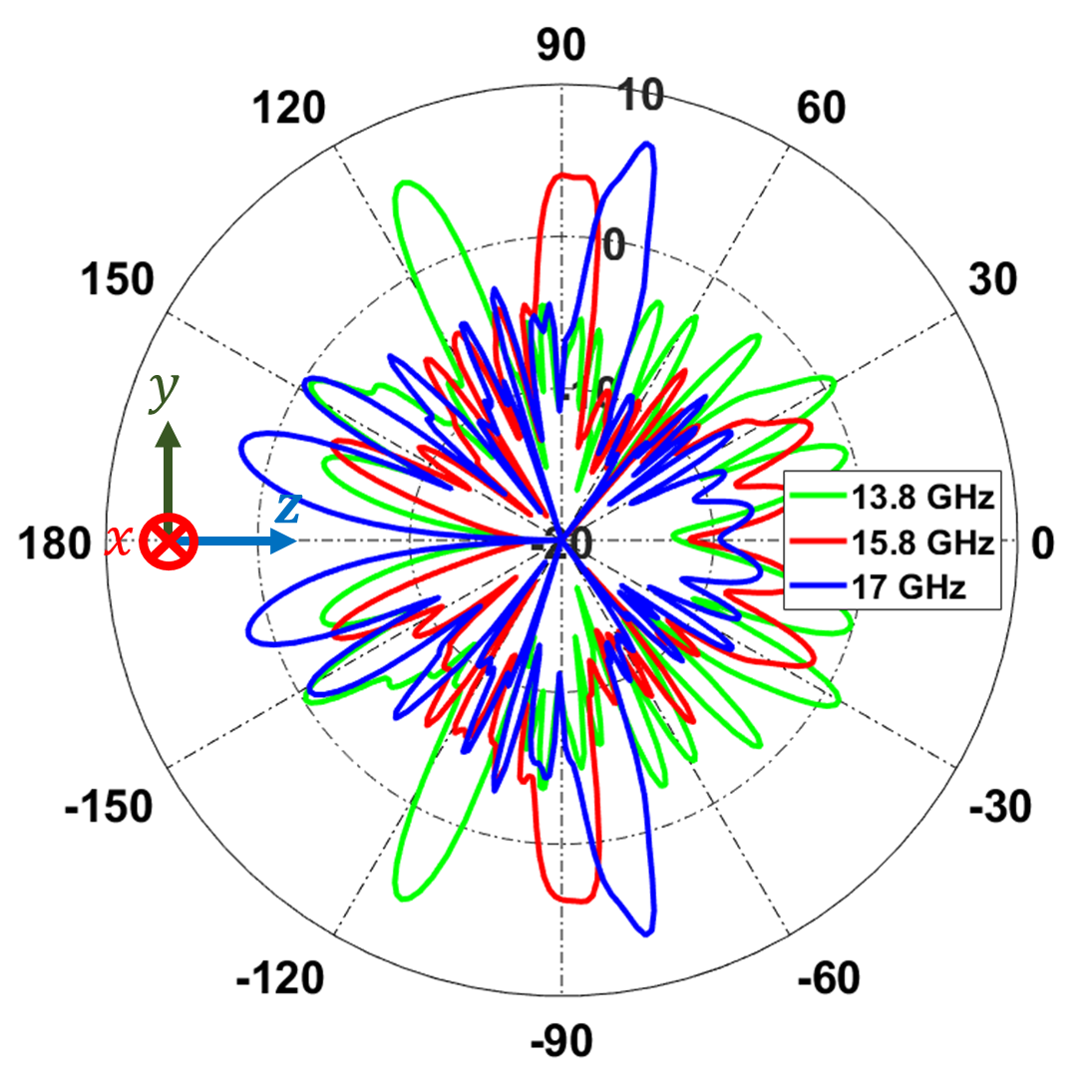}
	\label{Scan}
	}
	\end{center}
\caption{\subref{ArmAntPatPort1}The realized gain of the antenna in 3D display at 15.8 GHz, where it radiates in broadside and is excited by port 1.\subref{Scan}The realized gain of the antenna in the $\phi=90^{\circ}$ at the frequencies of 13.8 GHz, 15.8 GHz, and 17 GHz depicts the structure's scan limit.}
\label{ArmAntPatPort1-}
\end{figure}

The realized gain pattern at 15.8 GHz, where radiation occurs in a broadside direction, is shown in Fig. \ref{ArmAntPatPort1}. A comparison between the patterns of the zigzag arrangement (Fig. \ref{zigzag}) and the armchair arrangement reveals that the armchair structure is more suitable. 


As illustrated in Fig.\ref{S_Para}, the insertion loss decreases rapidly in the green area, where radiation occurs from the zigzag section of the structure. Fig.\ref{Ribbon} demonstrates that the phase velocity of the zigzag portion is negative, while that of the armchair portion is positive. Consequently, the zigzag section generates an additional beam in the backward direction, as observed in Fig.\ref{Scan} at the 17 GHz frequency. Due to this phenomenon, the frequencies within the green areas are undesirable for operation. Furthermore, as the frequency increases, the dispersion diagram in Fig.\ref{Hex_Ribbon_Armchair} indicates that the armchair section of the structure would exhibit two propagating modes in the fast wave region, with opposite phase velocities, leading to another additional backward beam.


The scanning limits of the structure are illustrated in Fig.\ref{Scan}, particularly highlighted in the yellow region of Fig. \ref{S_Para}, where the armchair section operates exclusively. Scanning starts at an angle of $114^{\circ}$ at a frequency of 13.8 GHz and, as it crosses the broadside, it reaches an angle of $78^{\circ}$ at 17 GHz.

At 15.8 GHz, the realized gain in the open stopband is less than -3 dB compared to the values observed at the ends of the scanning range. To the best of our knowledge, this unusual phenomenon does not occur in other unbalanced CRLH leaky-wave antennas. The 36-degree scan at both the top and bottom of the antenna surpasses the findings reported in \cite{abtahi2}, matching those in \cite{SinghAntenna}, even as it crosses the broadside.

\section*{Acknowledgment}
We would like to thank Prof. A. Zeidaabadi Nezhad form Isfahan University of Technology and Dr. R.J.B. Davis from University of California, San Diego for the fruitful and illuminating discussions.

\bibliographystyle{IEEEtran}
\bibliography{references1}
\newpage
\begin{IEEEbiography}[{\includegraphics[width=1in,height=1.25in,clip,keepaspectratio]{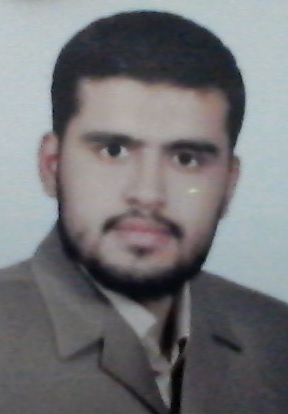}}]{SAYYED AHMAD ABTAHI} received the B.Sc. degree in electrical engineering from the Yazd University, Yazd, Iran, in 2014, and the M.Sc. degrees in telecommunications from Kashan University, Kashan, Iran, in 2018. He is currently
pursuing the Ph.D. degree in telecommunications with Isfahan University of Technology (IUT), Isfahan, Iran.  His research interests include topological metamaterials, leaky-wave antennas, periodic structures and photonic crystals.
\end{IEEEbiography}\hfil
\begin{IEEEbiography}[{\includegraphics[width=1in,height=1.25in,clip,keepaspectratio]{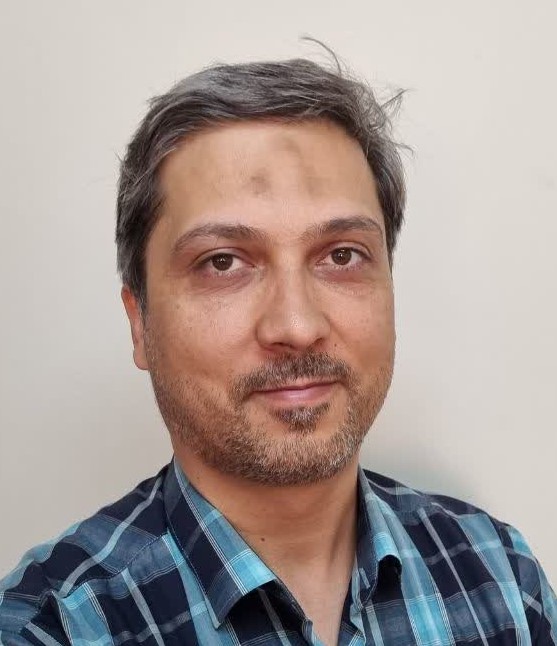}}]{MOHSEN  MADDAHALI} received the B.Sc. degree in electronics and telecommunications from the Isfahan University of Technology (IUT), Isfahan, Iran, in 2005, and the M.Sc. and Ph.D. degrees in electrical engineering from Tarbiat Modares University, Tehran, Iran, in 2008 and 2012, respectively.,Since 2012, he has been an Assistant Professor with the Department of Electrical and Computer Engineering, IUT. His research interests include computational electromagnetics, antenna theory, array antennas, new technologies in antenna design, and phased array antennas.
\end{IEEEbiography}
\begin{IEEEbiography}[{\includegraphics[width=1in,height=1.25in,clip,keepaspectratio]{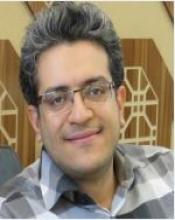}}]{AHMAD  BAKHTAFROUZ}received the B.S., M.S., and Ph.D. degrees from the Isfahan University of Technology (IUT), Isfahan, Iran, in 2006, 2009, and 2015, respectively, all in electrical engineering.,In 2016, he joined the faculty of the IUT, where he is currently an associate professor with the Electrical and Computer Engineering Department. His current research interests include millimeter-wave antennas, plasmonic devices, and periodic structures, such as electromagnetic bandgap structures (EBGs), artificial magnetic conductors (AMCs), and frequency selective surfaces (FSSs).
\end{IEEEbiography}

\EOD

\end{document}